\documentclass[lettersize,journal]{IEEEtran}
\usepackage{amsmath,amsfonts}
\usepackage{algorithmic}
\usepackage{algorithm}
\usepackage{array}
\usepackage[caption=false,font=normalsize,labelfont=sf,textfont=sf]{subfig}
\usepackage{textcomp}
\usepackage{stfloats}
\usepackage{url}
\usepackage{verbatim}
\usepackage{graphicx}
\usepackage{cite}
\usepackage{booktabs}
\usepackage{tabularx}
\usepackage{array}
\usepackage{tikz}
\usepackage{amsmath}
\usepackage{amssymb}
\usepackage[font=small,labelfont=bf,justification=centering]{caption}
\usetikzlibrary{arrows.meta, positioning, fit, backgrounds, shapes.geometric, calc}
\usepackage{xcolor}
\usepackage[most]{tcolorbox}
\usepackage{xcolor}
\usepackage{soul}
\usepackage{graphicx}
\usepackage{subcaption}
\usepackage{placeins}
\usepackage{multirow}
\usepackage{makecell}
\usepackage{adjustbox}
\usepackage{booktabs}
\usepackage{threeparttable}
\usepackage{multirow}
\usepackage[hidelinks]{hyperref} 
\usepackage{caption}

\newtcolorbox{rqsummarybox}{
colback=gray!10,
colframe=gray!45,
boxrule=0.5pt,
arc=1.5pt,
left=6pt,
right=6pt,
top=5pt,
bottom=5pt,
before skip=8pt,
after skip=8pt,
fontupper=\small
}

\begin{document}

\title{Understanding and Mitigating Representation Drift of Code Models under Software Distribution Shifts}

\author{Yiqing Shi, Zhifei Chen, Yangyang Zhao, Quanjun Zhang, Lin Chen and Wei Song,~\IEEEmembership{Senior Member,~IEEE}
\thanks{Yiqing Shi, Zhifei Chen, Quanjun Zhang, and Wei Song are with the School of Computer Science and Engineering, Nanjing University of Science and Technology, China (e-mail: 125106010860@njust.edu.cn; chenzhifei@njust.edu.cn;  quanjunzhang@njust.edu.cn;  wsong@njust.edu.cn).}
\thanks{Yangyang Zhao is with the School of Computer Science and Technology, Zhejiang Sci-Tech University, China (e-mail: yangyangzhao@zstu.edu.cn).}
\thanks{Lin Chen is with the School of Software, Nanjing University, China (e-mail: lchen@nju.edu.cn).}
\thanks{Zhifei Chen and Yangyang Zhao are the corresponding authors.}}

\markboth{IEEE Transactions on Software Engineering,~Vol.~52, No.~4, April~2026}%
{Shi \MakeLowercase{\textit{et al.}}: Understanding and Improving the Generalization of Code Representation Models}

\maketitle

\begin{abstract}
Deep learning-based code models often suffer performance degradation under software distribution shifts. Although recent studies have investigated the robustness of code models to out-of-distribution (OOD) data, how distribution shifts reshape internal representations and how such changes can guide adaptation decisions remain insufficiently understood. To address these gaps, we conduct a systematic empirical study following a diagnosis-to-adaptation workflow across seven tasks, five code models, and five distribution shift types.
Specifically, we measure generalization gaps between in-distribution (ID) and OOD data, evaluate task-adapted OOD detection signals, diagnose layer-wise representation geometry, and propose a representation-diagnostics-guided adaptation strategy. Our results show that: 1) performance degradation is decoupled from OOD detectability, with detectors achieving AUROC values below 59.00\% even when syntactic shifts reduce classification accuracy by up to 15.69\%; 2) settings with OOD performance degradation exhibit weaker cross-distribution alignment and more pronounced layer-wise geometric changes; and 3) diagnostics-guided data allocation and configuration search yield higher OOD scores than traditional adaptation strategies. Overall, our findings highlight the value of incorporating representation-level diagnostics into the evaluation and adaptation of code models under distribution shifts.
\end{abstract}


\section{Introduction}
\begin{figure}[!t]
    \centering
    \includegraphics[width=\linewidth]{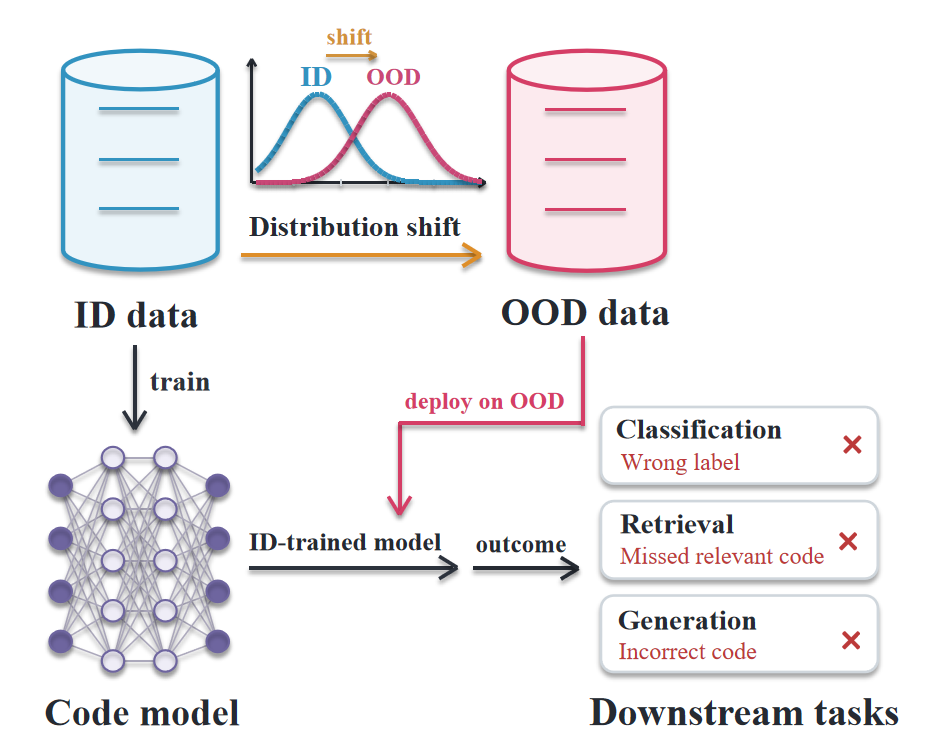}
    \caption{Illustration of code distribution shifts from training to inference.}
    \label{fig:distribution_shift}
\end{figure}

\IEEEPARstart{D}{eep} learning-based code representation models have been widely adopted in modern software engineering (SE) by learning lexical, syntactic, and semantic features from large-scale source code corpora.
Pretrained code models such as CodeBERT~\cite{feng2020codebert} and CodeT5~\cite{wang2021codet5}, which follow a pre-training and fine-tuning paradigm, have demonstrated strong performance across various SE tasks, including defect detection~\cite{zhou2019devign}, code search~\cite{gu2018deepcodesearch}, and program repair~\cite{chen2021sequencer}.
However, most existing evaluations implicitly assume that training and test data follow the same or similar distributions, an assumption that may not hold in real-world software development.
As illustrated in Fig.~\ref{fig:distribution_shift}, software systems continuously evolve due to version updates, API changes, developer turnover, library replacements, and evolving programming practices. As a result, code encountered during deployment may differ substantially from the data observed during training. 
This mismatch between training and deployment distributions is commonly referred to as \emph{distribution shift}~\cite{quinonero2008dataset}. 
In this context, samples following the training distribution are often referred to as \emph{in-distribution (ID)} samples, whereas samples whose characteristics deviate from the training distribution are regarded as \emph{out-of-distribution (OOD)} samples~\cite{yang2023generalized}. 
Such distribution shifts are becoming a critical challenge for building reliable code intelligence systems in practice~\cite{koh2021wilds}.

From the perspective of SE, distribution shifts can occur at different abstraction levels, including lexical changes in tokens and APIs, structural variations in program syntax and dependencies, and contextual changes across developers, software versions, and programming languages. Consequently, code models may encounter OOD data through both surface-form changes and deeper structural or contextual variations. 
Recent studies have begun to investigate the robustness of code representation models under distribution shifts~\cite{hu2023codes,shi2026codesplus,li2021uncertainty}. 
Hu et al.~\cite{hu2023codes} introduced CodeS to evaluate the generalization ability of code models under shifted scenarios beyond conventional ID settings. 
Shi et al.~\cite{shi2026codesplus} further extended this direction by considering more challenging structural variations and providing a broader assessment of code model robustness. 
These studies demonstrate that code models with excellent ID performance may still suffer considerable degradation when facing shifted code data. However, existing studies mainly characterize distribution shifts only through external model behavior, such as performance drops.

Beyond characterizing external performance drops, understanding and mitigating distribution-shift failures require both reliable OOD detection and effective model adaptation. For OOD detection, traditional methods flag samples that deviate from the training distribution using prediction confidence~\cite{hendrycks2017baseline} or representation distance~\cite{lee2018mahalanobis}. However, their reliability across heterogeneous tasks, which vary widely in output structures (e.g., classification logits vs.\ structured sequence generation), remains largely unverified. More fundamentally, existing evaluations treat code models as black boxes, offering limited insight into how distribution shifts and task-specific fine-tuning reshape the internal representation space, or how such internal drift relates to observable OOD failures.

To mitigate OOD failures, traditional adaptation strategies, including transfer learning and continual learning, have been explored to recover performance on shifted distributions while curbing catastrophic forgetting~\cite{pan2010survey,parisi2019continual,delange2022continual,robins1995catastrophic}. Yet, these methods typically apply a heuristic adaptation policy across all shifted samples and configurations, without diagnosing the underlying representation drift. Consequently, practitioners lack principled guidance on when adaptation can succeed without sacrificing ID capabilities, and how representation-level signals can be leveraged to direct adaptation decisions.

To bridge these gaps, we conduct a comprehensive empirical study of code representation models under distribution shifts. Unlike prior work that evaluates robustness solely from the perspective of external task degradation, we investigate the problem through a rigorous \textit{diagnosis-to-adaptation} workflow. Our study covers five representative code models, seven downstream SE tasks, and five distribution shift types. We first establish controlled ID/OOD evaluation settings to investigate how code representation models perform under different types of code distribution shifts (RQ1), evaluate whether task-adapted OOD detection signals can be used to identify shifted samples (RQ2), and characterize how distribution shifts affect the layer-wise representation space of code models (RQ3). We further evaluate the effectiveness of traditional adaptation methods in mitigating OOD failures (RQ4). Motivated by the observed association between internal representation changes and performance degradation caused by OOD in RQ3, we propose a representation-diagnostics-guided adaptation strategy that leverages geometric difficulty to direct sample allocation and configuration search (RQ5).

Through our empirical study, we obtain three key empirical findings: \textbf{1) Severe task performance degradation does not imply high OOD detectability.} For instance, while CST shift degrades classification accuracy by 12.62\%--15.69\%, all OOD detectors yield AUROC values below 59.00\%. This decoupling demonstrates that relying on existing OOD detectors to intercept deployment failures in code intelligence systems is unreliable, necessitating task-adapted detection signals.
\textbf{2) Representation changes are closely associated with OOD performance degradation.} Settings with performance declines in downstream tasks exhibit weaker ID-OOD Alignment (0.940 vs. 0.982) and more pronounced geometric changes than other settings. Layer-wise analysis further reveals heterogeneous representation changes across network depth and model architectures, suggesting a relationship between internal representation behavior and OOD performance loss.
\textbf{3) Representation diagnostics can guide more effective adaptation under distribution shifts.} Traditional adaptation methods exhibit varying trade-offs between OOD improvement and ID retention. By incorporating representation diagnostics into data allocation and configuration search, our approach achieves higher reported OOD scores than standard Replay across all five retrieval models, while maintaining comparable ID performance in these settings.

The main contributions of this paper are listed as follows:
\begin{itemize}
    \item \textbf{A comprehensive cross-task and cross-shift empirical benchmark.} We establish a unified ID/OOD evaluation framework spanning ten datasets, seven SE tasks, five representative code models, and five shift types, systematically uncovering how downstream task objectives interact with code distribution shifts.
    \item \textbf{A representation diagnostic framework for code models.} We introduce a multi-dimensional diagnostic suite to characterize representation changes under distribution shifts and their associations with downstream performance degradation.
    \item \textbf{A novel adaptation methodology.} We propose a representation-driven adaptation approach that translates internal geometric difficulty into sample allocation and configuration search policies, showing promising OOD improvements in selected settings while accounting for ID knowledge retention.
    \item \textbf{An open-source replication artifact.} We release our replication package\footnote{\url{https://github.com/Steven3199/Code-Model-Representation-Drift}} to facilitate future research on the robustness and adaptation of code representation models.
\end{itemize}

\textbf{Organization.} In Section \ref{sec:relatedwork}, we review the related work. In Section \ref{sec:design}, we go through our study design. Section \ref{sec:results} presents the results of our experiments. Section \ref{sec:discussion} discusses the potential
broader impact and future research directions, followed by threats to the validity of our study in Section \ref{sec:threats}. Finally, we conclude this work in Section \ref{sec:conclusion}.

\section{Related Work}\label{sec:relatedwork}
\subsection{Code Representation Models}
Code representation learning aims to encode source code into transferable representations for downstream SE tasks. Early approaches relied on explicit program structures: Code2vec~\cite{alon2019code2vec} and Code2seq~\cite{alon2019code2seq} represent programs using paths extracted from abstract syntax trees, while graph-based methods~\cite{allamanis2018learning} capture syntactic and semantic dependencies among program elements.

Large-scale Transformer pre-training has shifted code representation learning toward contextual representations from unlabeled corpora. RoBERTa~\cite{liu2019RoBERTa} provides a general-purpose baseline, while CodeBERT~\cite{feng2020codebert} extends pre-training to bimodal natural-language and programming-language data. GraphCodeBERT~\cite{guo2021graphcodebert} incorporates data-flow information to capture semantic dependencies beyond sequential tokens. For generation, PLBART~\cite{ahmad2021unified} adopts denoising sequence-to-sequence pre-training, while CodeT5~\cite{wang2021codet5} introduces identifier-aware objectives to capture the semantics of developer-assigned identifiers.

These code models have been widely applied to various SE tasks, including code understanding (e.g., defect detection~\cite{zhou2019devign} and clone detection~\cite{white2016clone}), code search~\cite{gu2018deepcodesearch}, program repair~\cite{chen2021sequencer}, code generation~\cite{chen2021evaluating}, and code translation~\cite{lu2021codexglue,khan2024xcodeeval}. These applications demonstrate the broad potential of pretrained code models in supporting diverse SE tasks.

\subsection{Evaluation of Code Representation Models}
The evaluation of code representation models has expanded from task-specific performance to broader analyses of model capability. Standardized resources such as CodeSearchNet~\cite{husain2019codesearchnet}, CodeXGLUE~\cite{lu2021codexglue}, and CodeNet~\cite{puri2021codenet} provide common datasets, metrics, and protocols for comparing models across SE tasks. A systematic comparison of 19 pretrained code models on 13 tasks~\cite{niu2023empiricalcomparison} further examines how architecture, modality, pre-training objectives, and training data affect downstream performance.

Beyond fixed-task comparisons, recent studies have examined model generalization across tasks and languages. Niu et al.~\cite{niu2023crosscodebench} proposed CrossCodeBench to evaluate cross-task generalization. They separated training and evaluation tasks and tested transfer to unseen tasks. Khan et al.~\cite{khan2024xcodeeval} introduced XCodeEval for multilingual and multitask evaluation. The benchmark covers code understanding, generation, translation, and retrieval. Yan et al.~\cite{yan2024codescope} developed CodeScope to evaluate code models across tasks, languages, difficulty levels, and efficiency dimensions. These studies broaden the evaluation of code model capabilities, although only some explicitly examine generalization to unseen downstream tasks.

Representation-level studies provide a complementary view of model capability. Karmakar and Robbes~\cite{karmakar2021codeknowledge} introduced probing tasks for surface, syntactic, structural, and semantic code properties, and later expanded this analysis through the INSPECT framework~\cite{karmakar2024inspect}. Troshin and Chirkova~\cite{troshin2022probing} similarly investigated layer-wise representations to examine whether code models encode syntax, identifiers, data flow, and semantic equivalence. Shi et al.~\cite{shi2023efficient} further compared pretrained and fine-tuned representations, showing that fine-tuning preserves much of the knowledge in the lower and intermediate layers while inducing larger changes near the upper layers. Together, these studies linked downstream performance to learned code representations but primarily focused on conventional evaluation settings.
In contrast, our study examines how distribution shifts affect both downstream performance and internal representations, linking representation-level changes to OOD detection and adaptation across diverse SE tasks.

\subsection{Distribution Shift in Software Engineering}
Code distribution shift occurs when the downstream task remains unchanged but the code distribution differs between training and deployment. CodeS~\cite{hu2023codes} defines several code-specific distribution shifts. \textit{User Shift} separates programs written by different developers, \textit{Time Shift} separates programs created in different periods, \textit{Token Shift} changes the frequency distribution of lexical units such as identifiers, and \textit{CST Shift} changes the syntactic structures represented by concrete syntax trees. Prior work also considers \textit{Language Shift} by evaluating code models across different programming languages~\cite{khan2024xcodeeval}. Software evolution may further induce Time Shift through changes in APIs and library interfaces~\cite{weyssow2023continual,liu2024codeupdatearena}.
Accordingly, we consider five types of distribution shift in this study: CST, Token, User, Time, and Language shifts, with API evolution treated as a form of temporal distribution change.

Several benchmarks and evaluation protocols have been developed to study these shifts. Li et al.~\cite{li2021uncertainty} examined project-, programmer-, and time-based shifts in code summarization and completion, while Nie et al.~\cite{nie2022evaluation} compared mixed-project, cross-project, and time-segmented evaluation for code summarization. CodeS~\cite{hu2023codes} provided User, Time, Token, and CST shifts for Java and Python classification, and CodeS+~\cite{shi2026codesplus} extends this setting with finer-grained structural variations. Arakelyan et al.~\cite{arakelyan2023distributional} further defined code domains using organizational, project, and module boundaries for summarization and generation. To mitigate distribution-shift effects, Weyssow et al.~\cite{weyssow2023continual} studied continual adaptation to evolving program streams containing previously unseen APIs, while Liu et al.~\cite{liu2024codeupdatearena} introduced CodeUpdateArena to evaluate model updating as APIs evolve over time. These studies reveal important robustness challenges under software evolution, but typically focus on specific tasks, shifts, or adaptation settings. 
Our study provides a unified analysis of code distribution shift by connecting performance degradation, representation drift, and adaptation across multiple tasks, shifts, and models.

\section{Study Design}\label{sec:design}
\subsection{Research Questions}
The goal of this study is to understand not only whether code representation models fail under distribution shifts, but also how such failures can be diagnosed and whether diagnostic information can support more effective adaptation. To address these objectives, we formulate five research questions.

\textbf{RQ1: How do code representation models perform under different types of code distribution shifts across diverse SE scenarios?}
Existing benchmarks have demonstrated the vulnerability of code models to distribution shifts~\cite{hu2023codes,shi2026codesplus,arakelyan2023distributional}. However, it remains unclear whether the observed degradation follows consistent patterns across different downstream tasks, model architectures, and shift types. A systematic evaluation across these dimensions is therefore needed to characterize the robustness of code representation models.

\textbf{RQ2: How effective are task-adapted OOD detection signals for identifying shifted code samples?}
Identifying shifted samples is important for reliable deployment, but existing OOD detection signals may not be equally effective across different SE tasks. In particular, classification, retrieval, and generation produce different prediction outputs, which may affect the effectiveness of detection signals. This motivates investigating task-adapted OOD detection methods across diverse downstream scenarios.

\textbf{RQ3: How do distribution shifts affect the representation space of code models?}
Task-level performance reveals whether a model succeeds or fails, but provides limited insight into how distribution shifts affect its internal representations. Moreover, representation invariance alone does not guarantee successful generalization across domains~\cite{bendavid2006neurips,zhao2019invariant}. Understanding representation changes, particularly those induced by task-specific fine-tuning, is therefore important for explaining model behavior under distribution shifts.

\textbf{RQ4: Can existing adaptation methods mitigate performance degradation caused by code distribution shifts?}
Adapting models with shifted-distribution data provides a potential means of mitigating performance degradation. However, adaptation may improve OOD performance at the expense of ID capability, reflecting the stability--plasticity trade-off studied in continual learning~\cite{lu2025stability}. This motivates evaluating existing adaptation strategies in terms of both OOD improvement and ID retention.

\textbf{RQ5: Can representation diagnostics guide more effective adaptation under code distribution shift?}
Prior studies have explored difficulty-aware target sample selection in domain adaptation, showing the potential value of distinguishing samples with different adaptation difficulties~\cite{zuo2021challenging,lifshitz2021sample}. This motivates investigating whether representation diagnostics can provide useful signals for estimating the relative difficulty of OOD data and guiding adaptation configuration selection in code models.

\subsection{Overall Workflow and Framework}
\begin{figure*}[t]
\centering
\includegraphics[width=0.9\textwidth]{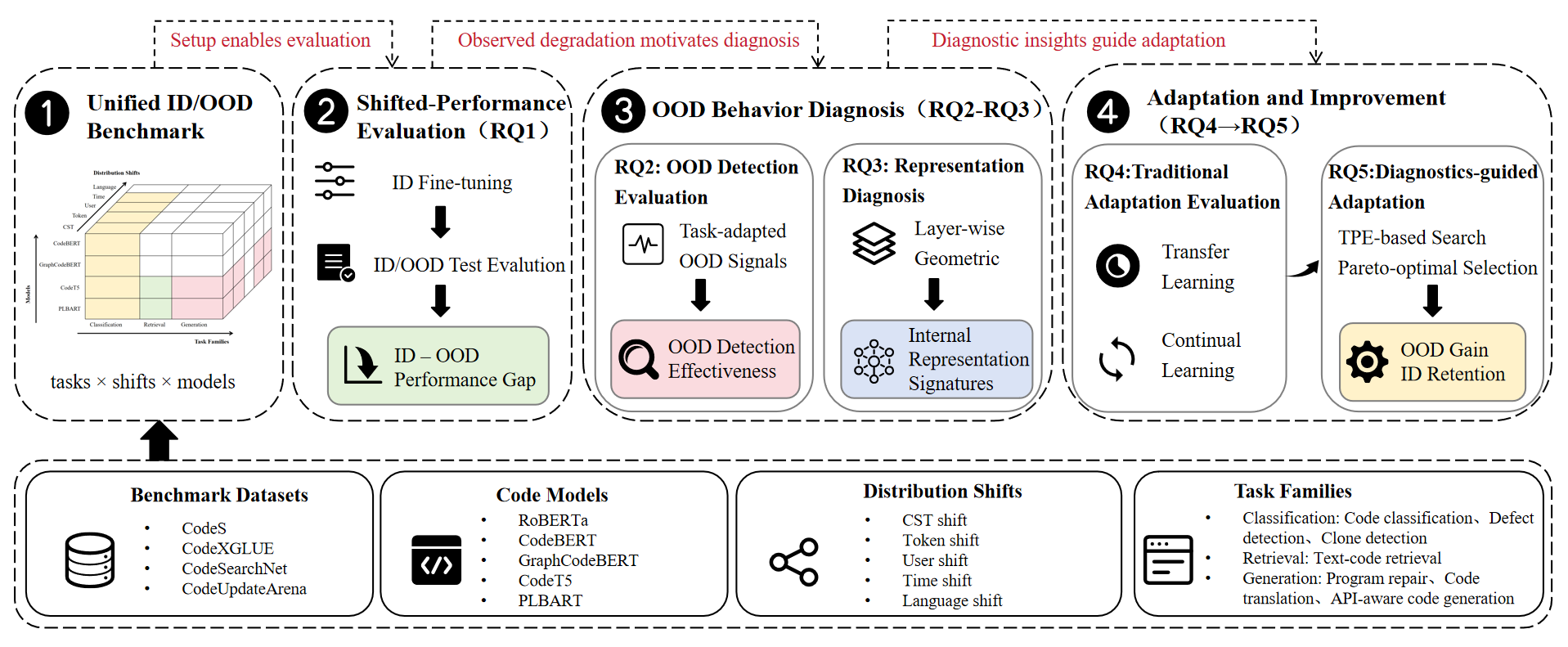}
\caption{Overview of the diagnosis-to-adaptation workflow: unified ID/OOD setup, shifted-performance evaluation (RQ1), OOD behavior diagnosis (RQ2--RQ3), adaptation and improvement (RQ4--RQ5).}
\label{fig:study_design}
\end{figure*}

To systematically investigate the above research questions, we organize our study into a four-stage \textit{diagnosis-to-adaptation} workflow, as illustrated in Figure~\ref{fig:study_design}. The stages progressively move from external behavioral evaluation to internal representation diagnosis and targeted adaptation:

\textbf{Stage 1: Unified ID/OOD Benchmark Setup.}
We first establish controlled ID/OOD evaluation settings by collecting representative code datasets and defining shifted test distributions under different types of code distribution shifts.
This stage provides the experimental foundation for comparing model behavior across diverse downstream SE scenarios.

\textbf{Stage 2: Shifted-Performance Evaluation.}
Based on the constructed ID/OOD settings, we fine-tune representative code representation models on ID training data and evaluate their performance on both ID and OOD test sets.
This stage addresses \textbf{RQ1} by examining how model performance changes when the test distribution differs from the training distribution across different models, tasks, and shift types.

\textbf{Stage 3: OOD Behavior Diagnosis.}
We then diagnose OOD behavior from two complementary perspectives.
First, we evaluate whether shifted code samples can be identified by task-adapted OOD detection signals, corresponding to \textbf{RQ2}.
Second, we move beyond task-level performance and analyze how distribution shifts and task-specific fine-tuning affect the internal representation space of code models, corresponding to \textbf{RQ3}.
This stage is designed to connect observable OOD behavior with internal representation changes.

\textbf{Stage 4: Adaptation and Improvement.}
Finally, we study whether model adaptation can mitigate failures under shifted distributions.
We first evaluate standard transfer learning and continual learning methods, corresponding to \textbf{RQ4}.
We then investigate whether representation diagnostics can provide useful signals for guiding adaptation decisions, corresponding to \textbf{RQ5}.
This stage further connects representation-level diagnosis with practical adaptation.

\subsection{Subject Models}
We select five representative code representation models: RoBERTa~\cite{liu2019RoBERTa}, CodeBERT~\cite{feng2020codebert}, GraphCodeBERT~\cite{guo2021graphcodebert}, CodeT5~\cite{wang2021codet5}, and PLBART~\cite{ahmad2021unified}. These model families have been widely evaluated across code understanding and generation tasks in established SE benchmarks, surveys, and systematic empirical comparisons~\cite{lu2021codexglue,niu2022survey,niu2023empiricalcomparison}. They also cover complementary model architectures and pre-training strategies. RoBERTa serves as a general-purpose encoder baseline; CodeBERT and GraphCodeBERT represent code-specific sequential and structure-aware encoders, respectively; and CodeT5 and PLBART represent encoder--decoder models designed for both understanding and generation. This selection enables us to compare general-purpose and code-specific pretraining, sequential and structure-aware representations, and encoder-only and encoder--decoder architectures under the same distribution-shift settings.

\subsection{Datasets and Distribution Shifts}
\label{sec:datasets}
Table~\ref{tab:dataset_shift_task} summarizes the datasets, distribution shifts, downstream tasks, and evaluation metrics used in this study. We retain the original training and test splits provided by each benchmark as the ID training and ID test sets, respectively. For each dataset--shift setting, the corresponding OOD test set contains the same number of samples as the ID test set. The third and fourth columns report the sizes of the ID training set and each of the ID and OOD test sets, respectively. Overall, our study covers ten code-specific datasets, seven downstream tasks spanning classification, retrieval, and generation, and five types of distribution shift, including lexical and structural changes, developer and temporal differences, cross-language variation, and API evolution.

We retain benchmark-defined shift settings when available. Specifically, we use the CST, Token, User, and Time shifts samples from CodeS~\cite{hu2023codes}, the Language Shift samples in two translation directions from CodeXGLUE-CodeTrans~\cite{lu2021codexglue}, and an API-evolution setting constructed based on CodeUpdateArena~\cite{liu2024codeupdatearena}. In addition, to support comparisons of shift effects between different downstream tasks, we construct Token and CST shifts for code samples from CodeXGLUE-Devign, CodeXGLUE-BigCloneBench, CodeSearchNet, and CodeXGLUE-Bugs2Fix. This choice is motivated by prior evidence that Token and CST shifts have a greater impact on code models~\cite{hu2023codes}. To implement these shifts, we apply semantics-preserving NatGen transformations~\cite{chakraborty2022natgen} to Java and Python programs, and also implement the corresponding C transformations over concrete syntax trees generated by Tree-sitter~\cite{treesitter}. Through these transformations, we construct a paired OOD sample for each original test sample in these datasets.

\begin{table*}[t]
\centering
\caption{DATASETS AND METRICS}
\label{tab:dataset_shift_task}
\scriptsize
\setlength{\tabcolsep}{4pt}
\renewcommand{\arraystretch}{1.35}

\begin{tabular}{
    >{\centering\arraybackslash}p{0.60cm}
    >{\raggedright\arraybackslash}p{3.15cm}
    >{\raggedright\arraybackslash}p{1.50cm}
    >{\raggedright\arraybackslash}p{1.20cm}
    >{\raggedright\arraybackslash}p{1.30cm}
    >{\raggedright\arraybackslash}p{2.00cm}
    >{\raggedright\arraybackslash}p{2.10cm}
    >{\raggedright\arraybackslash}p{4.10cm}
}
\toprule
\textbf{ID}
& \textbf{Dataset}
& \textbf{\#Training\_Set}
& \textbf{\#Test\_Set}
& \textbf{Language}
& \textbf{Shift Type}
& \textbf{Downstream Task}
& \textbf{Evaluation Metrics}
\\
\midrule

D1
& CodeS-Java250
& 45,000
& 15,000
& Java
& CST, Token, User
& Code classification
& Accuracy, Macro-F1, Top-5 Accuracy
\\

D2
& CodeS-Python75
& 54,900
& 10,050
& Python
& CST, Token, User, Time
& Code classification
& Accuracy, Macro-F1, Top-5 Accuracy
\\

D3
& CodeS-Python800
& 144,000
& 48,000
& Python
& CST, Token, User
& Code classification
& Accuracy, Macro-F1, Top-5 Accuracy
\\

D4
& CodeXGLUE-Devign
& 21,854
& 2,732
& C
& CST, Token
& Defect detection
& Precision, Recall, F1-score
\\

D5
& CodeXGLUE-BigCloneBench
& 901,028
& 415,416
& Java
& CST, Token
& Clone detection
& Precision, Recall, F1-score
\\

D6
& CodeSearchNet
& 412,178
& 22,176
& Java, Python
& CST, Token
& Text-code retrieval
& MRR, Recall@1, Recall@5
\\

D7
& CodeXGLUE-Bugs2Fix-small
& 46,680
& 5,835
& Java
& CST, Token
& Program repair
& Exact Match, BLEU, CodeBLEU
\\

D8
& CodeXGLUE-Bugs2Fix-medium
& 52,364
& 6,545
& Java
& CST, Token
& Program repair
& Exact Match, BLEU, CodeBLEU
\\

D9
& CodeXGLUE-CodeTrans
& 10,295
& 1,000
& Java, C\#
& Language
& Code translation
& Exact Match, BLEU, CodeBLEU
\\

D10
& CodeUpdateArena
& 536
& 134
& Python
& Time
& API-aware code generation
& Syntax Pass, Executability, Pass@1
\\

\bottomrule
\end{tabular}
\end{table*}

\subsection{Analysis Method for RQ1}
RQ1 evaluates how code representation models generalize under different code distribution shifts across classification, retrieval, and generation tasks. We use task-specific metrics that are commonly adopted in code-model evaluation~\cite{niu2022survey}.

The classification tasks include code classification, defect detection, and clone detection. For multiclass code classification, we report Accuracy, Macro-F1, and Top-5 Accuracy~\cite{niu2022survey}, which respectively capture overall correctness, class-balanced performance, and top-$k$ prediction quality. For defect and clone detection, we report Precision, Recall, and F1-score following established SE evaluation protocols~\cite{lu2021codexglue}.

For the text-code retrieval task, we report Mean Reciprocal Rank (MRR), Recall@1, and Recall@5 following established code-retrieval evaluation protocols~\cite{husain2019codesearchnet}. These metrics characterize the ranking position of relevant code and retrieval effectiveness at different cutoffs.

For program repair and code translation, we report Exact Match, BLEU~\cite{papineni2002bleu}, and CodeBLEU~\cite{ren2020codebleu}, following commonly used code-generation evaluation protocols~\cite{lu2021codexglue}. For API-aware code generation, we report Syntax Pass, Executability, and Pass@1.
Syntax Pass and Executability are used to assess syntactic validity and execution validity, respectively, while Pass@1~\cite{chen2021evaluating} measures functional correctness following established
code-generation evaluation practices.

For each model--task--shift configuration, the model is fine-tuned exclusively on the ID training set and evaluated separately on the ID and OOD test sets. We measure robustness via the performance gap between ID and OOD evaluations. A larger drop indicates greater vulnerability to the shift, whereas comparable or higher OOD scores indicate strong cross-distribution generalization.

\subsection{Analysis Method for RQ2}
RQ2 examines whether shifted code samples can be distinguished from ID samples using OOD signals suited to the prediction structure of each task. OOD detection aims to identify whether a test sample follows the training distribution. 
In general, OOD detectors distinguish shifted samples by exploiting the observation that OOD inputs often lead to lower prediction confidence, higher uncertainty, or larger distances from the learned ID representation distribution.
Since different tasks in our study produce different prediction structures, we investigate task-specific detection signals rather than applying a unified detector across all tasks.

For classification tasks which directly produce logits and class probabilities, we apply nine established OOD detectors: MSP~\cite{hendrycks2017baseline}, predictive Entropy~\cite{kirsch2021pitfalls}, MaxLogit~\cite{hendrycks2022scaling}, Energy~\cite{liu2020energy}, Mahalanobis distance~\cite{lee2018mahalanobis}, Mahalanobis++~\cite{muller2025mahalanobis}, KNN~\cite{sun2022knn}, ViM~\cite{wang2022vim}, and GradNorm~\cite{huang2021gradnorm}. These detectors characterize OOD samples from complementary perspectives, including confidence estimation, uncertainty measurement, and representation-space deviation.

Retrieval and generation tasks do not produce fixed class-probability outputs, thus we construct three task-adapted detection signals for each task. 
For the text--code retrieval task, Matching Score~\cite{husain2019codesearchnet} uses the highest query--code relevance score, following the relevance estimation used in semantic code retrieval. We further define Ranking Entropy~\cite{kirsch2021pitfalls} based on the probability distribution over candidate code samples to characterize the uncertainty in retrieval ranking. KNN Distance~\cite{sun2022knn} measures representation-space deviation from ID samples following nearest-neighbor-based OOD detection.
For the code generation task, Seq. NLL measures the negative log-likelihood assigned to the target sequence, and we include this signal to characterize model confidence under distribution shifts~\cite{hajipour2024simscood}. Token Entropy~\cite{kirsch2021pitfalls} captures predictive uncertainty during decoding, and Encoder Distance~\cite{lee2018mahalanobis,wang2022vim} measures deviation between input representations and the ID representation space following feature-based OOD detection methods.

We evaluate the effectiveness of all selected OOD detectors and signals using AUROC, AUPR, and TNR@95, which are commonly adopted metrics in OOD detection evaluation~\cite{hendrycks2017baseline,hendrycks2022scaling}. AUROC and AUPR quantify the separability between ID and OOD samples across different decision thresholds, while TNR@95 measures the proportion of OOD samples correctly rejected when 95\% of ID samples are retained~\cite{hendrycks2017baseline}.

\subsection{Analysis Method for RQ3}
RQ3 investigates how distribution shifts affect the representation space of code models. For each model--task--shift setting, we extract layer-wise hidden representations from both ID and OOD test samples using fine-tuned models. Let $h_{i,l}^{S} \in \mathbb{R}^{d_{l}}$ denote the pooled representation of sample $i$ from distribution $S$ at layer $l$, where $S \in \{\mathrm{ID}, \mathrm{OOD}\}$.

Since models produce token-level hidden states, we apply pooling to obtain sample-level representations. RoBERTa, CodeBERT, and GraphCodeBERT use the first special token representation, while CodeT5 and PLBART use mean pooling over non-padding encoder tokens. For encoder--decoder models, encoder states are used for Alignment, Anisotropy, and Effective Rank, whereas decoder states are used for generation-related Entropy.

We then characterize the representation space using four complementary diagnostics: Alignment, Anisotropy, Effective Rank, and Entropy. We compare their characteristics between ID and OOD distributions to examine how distribution shifts reshape internal model representations:

\textbf{Alignment}~\cite{wang2020understanding}  measures the consistency between ID and OOD representations inspired by alignment in contrastive learning. 
Let $\mathcal{U}$ denote the set of matched units shared between ID and OOD data. For paired shifts such as Token and CST, each shared sample index defines a unit, whereas for non-paired shifts such as User and Time, shared class labels define units represented by class centroids. Let $\mu_{u,l}^{D}$ denote the corresponding sample representation or class centroid at layer $l$ under distribution $D$. The layer-wise Alignment score is defined as:
\begin{equation}
\mathrm{Align}_{l} = \frac{1}{|\mathcal{U}|}\sum_{u\in\mathcal{U}}\cos(\mu_{u,l}^{\mathrm{ID}},\mu_{u,l}^{\mathrm{OOD}})
\end{equation}
Higher values indicate greater consistency between ID and OOD representations.

\textbf{Anisotropy}~\cite{ethayarajh2019contextual} measures the directional concentration of representations in the embedding space. We compute the average pairwise cosine similarity among $L_{2}$-normalized representations separately for ID and OOD samples at each layer, excluding self-similarities. Let $N_{S}$ denote the number of samples from distribution $S \in \{\mathrm{ID}, \mathrm{OOD}\}$, and $h_{i,l}^{S,m}$ the corresponding pooled representation for model $m$. We define:
\begin{equation}
\mathrm{Aniso}_{l}^{S,m} = \frac{2}{N_{S}(N_{S}-1)}\sum_{i<j}\cos(h_{i,l}^{S,m},h_{j,l}^{S,m})
\end{equation}
Higher values indicate stronger directional concentration, suggesting that representations are less dispersed and potentially less discriminative.

\textbf{Effective Rank}~\cite{roy2007effective} estimates the effective dimensionality of the representation space. We center the representation matrix for each distribution and layer and apply singular value decomposition. Let $\sigma^{S,m}_{l,j}$ denote the $j$-th singular value at layer $l$, with $r$ singular values in total. The effective rank is
\begin{equation}
\mathrm{ERank}^{S,m}_{l}=\exp\!\left(-\sum_{j=1}^{r}p^{S,m}_{l,j}\log p^{S,m}_{l,j}\right),
\label{eq:rq3-erank}
\end{equation}
where
\begin{equation}
 p^{S,m}_{l,j}=\frac{\sigma^{S,m}_{l,j}}{\sum_{k=1}^{r}\sigma^{S,m}_{l,k}}.
\label{eq:rq3-rank-prob}
\end{equation}
Higher effective rank indicates a more distributed singular-value spectrum, while lower effective rank indicates that variance is concentrated in fewer directions.

\textbf{Entropy}~\cite{kirsch2021pitfalls} measures the predictive uncertainty of model outputs induced by representations from different distributions.
We apply the fixed fine-tuned task-specific output mapping to ID and OOD representations at each layer, with task-specific calculations. For classification, entropy is computed from class probability distributions.
For retrieval, we calculate similarities between the same ID queries and the ID/OOD code banks separately, followed by temperature scaling and softmax to obtain probability distributions. For generation, we evaluate the ID and OOD source--target pairs separately under teacher forcing, project decoder hidden states onto the vocabulary space, and compute token-level entropy over valid target positions. We then average sample-level entropy values to obtain the layer-wise diagnostic. Higher entropy indicates greater uncertainty under the corresponding task-specific calculation.

Alignment directly compares matched units across distributions, whereas Anisotropy, Effective Rank, and Entropy are calculated separately for ID and OOD data at each layer. 
To quantify ID--OOD representation changes, we compute paired differences for Anisotropy, Effective Rank, and Entropy at corresponding layers. For Anisotropy and Entropy, we define $\Delta=\mathrm{OOD}-\mathrm{ID}$, whereas for Effective Rank, $\Delta=\mathrm{ID}-\mathrm{OOD}$. Thus, positive differences indicate stronger directional concentration, reduced effective dimensionality, or increased predictive uncertainty under OOD conditions. 

We summarize these changes using the mean difference and the proportion of layer-wise records with $\Delta>0$. For each reported group, we apply the paired Wilcoxon signed-rank test to the ID-OOD differences, with Holm correction for multiple comparisons, and report the absolute rank-biserial correlation as the effect size. 

To further examine the effects of fine-tuning, we also compare pretrained and fine-tuned model representations in representative settings, using the same ID/OOD samples and representation extraction procedures.

\begin{figure*}[t]
    \centering
    \includegraphics[
        width=0.8\textwidth,
        keepaspectratio
    ]{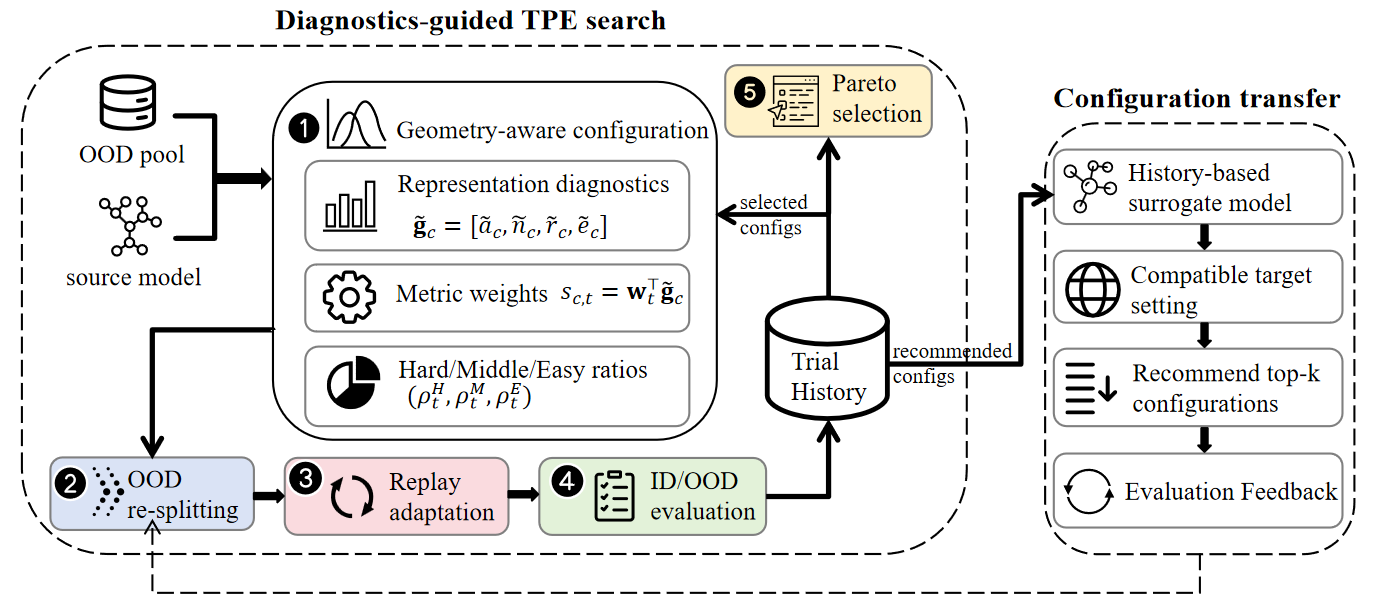}
    \caption{Overview of the representation-diagnostics-guided adaptation framework for RQ5.}
    \label{fig:rq5_framework}
\end{figure*}

\subsection{Analysis Method for RQ4}
RQ4 evaluates whether existing adaptation methods can improve OOD performance while preserving the model's behavior on the original ID distribution. 
Because exhaustively adapting all settings is computationally expensive and partly redundant, we select a representative subset of challenging dataset--shift settings for RQ4 and RQ5. 
Following prior work on representative benchmark subset discovery and redundancy reduction~\cite{subramani2025simba,zhan2022benchsubset}, we retain settings whose first quartile of relative OOD degradation is above zero, so that selected shifts reflect consistent degradation across most combinations. We then prioritize settings with larger median degradation within each task family and remove redundant task--shift combinations while preserving task and shift diversity.

We evaluate three representative adaptation strategies. \textbf{Transfer Learning (TL)}~\cite{pan2010survey} fine-tunes all trainable parameters on OOD data, enabling the model to adapt to the shifted distribution without explicit preservation of ID knowledge. \textbf{Elastic Weight Consolidation (EWC)}~\cite{kirkpatrick2017ewc} introduces parameter-level regularization based on parameter importance estimated from the ID task, reducing excessive updates to important parameters during adaptation. \textbf{Replay}~\cite{chaudhry2019episodic,robins1995catastrophic}
jointly optimizes on OOD samples and retained ID examples, providing a direct mechanism to balance adaptation and ID knowledge preservation.

For each selected setting, the OOD data pool is split into an adaptation subset (80\%) and a disjoint evaluation subset (20\%). Only the OOD adaptation subset is used for model updates. We then evaluate each adapted model on the original ID test set and the held-out OOD evaluation subset. The resulting OOD and ID scores are used to assess adaptation effectiveness and retention of the original capabilities, respectively.

\subsection{Analysis Method for RQ5}

RQ5 investigates whether the final-layer representation diagnostics developed in RQ3 can guide the selection of adaptation configurations under code distribution shift. Whereas RQ4 compares predefined adaptation methods, RQ5 uses the same representative settings to search for configurations of Replay-based adaptation. Replay is retained as the underlying adaptation mechanism because it incorporates ID examples alongside OOD data. 

Our design is inspired by difficulty-aware sample selection in domain adaptation~\cite{zuo2021challenging,lifshitz2021sample}. Rather than treating all OOD groups identically, it uses representation scores as indicators of their relative adaptation difficulty and changes how their samples are allocated between adaptation and validation.
As illustrated in Figure~\ref{fig:rq5_framework}, the method contains the following two modules.

\textbf{Module 1: Diagnostics-guided TPE (Tree-structured Parzen Estimator) search.}
The first module iteratively searches adaptation configurations using TPE~\cite{bergstra2011tpe} and evaluates them through the following five steps.

\textit{Step 1: Geometry-aware configuration.}
Given a fixed source model and OOD pool, TPE searches two groups of variables: the weights assigned to the four representation diagnostics and the allocation preferences $(\rho_t^{H},\rho_t^{M},\rho_t^{E})$ for hard, middle, and easy OOD classes. 
The diagnostic definitions, global OOD train--test ratio, and Replay hyperparameters remain fixed throughout the search.

For each OOD class $c$, the four diagnostics extracted from the final representation layer are min--max normalized and direction-adjusted, with higher values consistently indicating greater difficulty:
\begin{equation}
\tilde{\mathbf{g}}_c =
[\tilde{a}_c,\tilde{n}_c,\tilde{r}_c,\tilde{e}_c].
\label{eq:diagnostic_vector}
\end{equation}
corresponding to Alignment, Anisotropy, Effective Rank, and Entropy, respectively.

For trial $t$, TPE proposes a diagnostic-weight vector $\mathbf{w}_t$ and computes the corresponding difficulty score:
\begin{equation}
s_{c,t} =
\mathbf{w}_t^\top \tilde{\mathbf{g}}_c.
\label{eq:difficulty_score}
\end{equation}

\emph{Step 2: OOD re-splitting.}
We use the same dataset--shift settings as in RQ4. For each setting, the OOD data pool is divided into adaptation and validation subsets with a fixed overall ratio of 80:20.

The OOD classes are ranked in descending order according to $s_{c,t}$ and divided into three groups: the top 25\% as hard, the middle 50\% as middle, and the bottom 25\% as easy. The allocation preferences $(\rho_t^H, \rho_t^M, \rho_t^E)$, subject to $\rho_t^H \leq \rho_t^M \leq \rho_t^E$, determine the relative validation allocation for each group.

Specifically, class-level validation allocations are proportional to the product of class sample counts and their corresponding group preferences. Harder classes therefore contribute relatively more samples to adaptation, whereas easier classes contribute relatively more samples to validation.

\textit{Step 3: Replay adaptation.}
Each candidate starts from the same checkpoint obtained in RQ1. An ID replay buffer is constructed from the ID training data, and the model is updated jointly using OOD and retained ID samples according to
\begin{equation}
\mathcal{L}_{\mathrm{Replay}}=\mathcal{L}_{\mathrm{OOD}}+\lambda\mathcal{L}_{\mathrm{ID}},
\label{eq:rq5-replay}
\end{equation}
where $\lambda$ (1 by default) and the remaining Replay settings are fixed across trials.

\textit{Step 4: ID/OOD evaluation.}
The adapted model is evaluated on the original ID test set and the trial-specific OOD validation subset. The OOD performance, together with the corresponding configuration, provides feedback for the subsequent search, while ID performance is recorded to assess retention. 

\textit{Step 5: Pareto selection.}
Pareto selection~\cite{deb2002nsga2} identifies non-dominated configurations based on selected task-relevant OOD validation performance objectives. ID performance is monitored separately to assess potential retention effects, but is not included in the optimization objectives. The resulting candidates guide subsequent TPE proposals. Steps 1--5 are repeated until the predefined search budget is exhausted.

\textbf{Module 2: Configuration transfer.}
The second module reuses the accumulated trial history to reduce search cost for compatible target settings. A surrogate model uses historical configurations and their outcomes to recommend the top-$k$ candidates for compatible target settings. These candidates are evaluated using the same procedure, and the resulting feedback informs subsequent configuration search. We restrict configuration transfer to settings with compatible model and task configurations, without transferring across different backbones or task families.

\subsection{Implementation Details}
Our experiments are implemented in Python 3.8.5 using TensorFlow 2.5.0, and are conducted on Ubuntu 20.04 equipped with a 4-core CPU, one NVIDIA TITAN RTX GPU with 24~GB of memory, and 64~GB of system memory. 

We use the publicly released checkpoints of RoBERTa, CodeBERT, GraphCodeBERT, CodeT5, and PLBART as backbone models. For each task, these models follow consistent preprocessing, input-length constraints, optimization settings, and checkpoint-selection procedures whenever applicable. We retain the official training and test partitions as the ID training and test sets, respectively, and obtain the corresponding OOD test sets using benchmark-defined shifts or the transformations described in Section~\ref{sec:datasets}.

\newcommand{\perfdrop}[1]{{\color{red}#1$\downarrow$}}
\newcommand{\perfgain}[1]{{\color{blue}#1$\uparrow$}}
\newcommand{\perfsame}[1]{#1}

\begin{table*}[t]
\centering
\caption{ID--OOD PERFORMANCE CHANGES UNDER DISTRIBUTION SHIFTS}
\label{tab:rq1_performance}

\footnotesize
\setlength{\tabcolsep}{0.45pt}
\renewcommand{\arraystretch}{1.10}

\resizebox{\textwidth}{!}{
\begin{tabular}{@{}c c ccc ccc ccc ccc ccc@{}}
\toprule

\textbf{Dataset}
&
\textbf{Shift}
&
\multicolumn{3}{c}{\textbf{RoBERTa}}
&
\multicolumn{3}{c}{\textbf{CodeBERT}}
&
\multicolumn{3}{c}{\textbf{GraphCodeBERT}}
&
\multicolumn{3}{c}{\textbf{CodeT5}}
&
\multicolumn{3}{c}{\textbf{PLBART}}
\\

\midrule

\multicolumn{2}{l}{\textbf{Code Classification}}
&
Acc. & Macro-F1 & Top-5
&
Acc. & Macro-F1 & Top-5
&
Acc. & Macro-F1 & Top-5
&
Acc. & Macro-F1 & Top-5
&
Acc. & Macro-F1 & Top-5
\\
\midrule

\multirow{3}{*}{D1}
& CST
& \perfdrop{12.27} & \perfdrop{11.66} & \perfdrop{10.29}
& \perfdrop{12.07} & \perfdrop{11.49} & \perfdrop{10.26}
& \perfdrop{6.40} & \perfdrop{6.00} & \perfdrop{5.22}
& \perfdrop{10.87} & \perfdrop{10.35} & \perfdrop{8.85}
& \perfdrop{8.33} & \perfdrop{8.02} & \perfdrop{6.57}
\\

& Token
& \perfdrop{10.99} & \perfdrop{10.82} & \perfdrop{7.44}
& \perfdrop{10.38} & \perfdrop{10.20} & \perfdrop{7.12}
& \perfdrop{6.31} & \perfdrop{6.13} & \perfdrop{4.11}
& \perfdrop{9.67} & \perfdrop{9.73} & \perfdrop{6.79}
& \perfdrop{9.34} & \perfdrop{9.21} & \perfdrop{5.77}
\\

& User
& \perfdrop{0.39} & \perfdrop{1.58} & \perfgain{0.28}
& \perfsame{0.00} & \perfdrop{0.97} & \perfgain{0.38}
& \perfdrop{0.28} & \perfdrop{0.24} & \perfdrop{0.05}
& \perfdrop{0.23} & \perfdrop{0.61} & \perfdrop{0.06}
& \perfdrop{0.08} & \perfdrop{0.36} & \perfgain{0.41}
\\

\addlinespace[1pt]

\multirow{4}{*}{D2}
& CST
& \perfdrop{15.69} & \perfdrop{14.59} & \perfdrop{13.44}
& \perfdrop{15.18} & \perfdrop{13.84} & \perfdrop{13.31}
& \perfdrop{14.98} & \perfdrop{13.82} & \perfdrop{13.21}
& \perfdrop{13.97} & \perfdrop{12.94} & \perfdrop{12.32}
& \perfdrop{12.62} & \perfdrop{11.60} & \perfdrop{10.79}
\\

& Token
& \perfdrop{1.88} & \perfdrop{1.97} & \perfdrop{0.02}
& \perfdrop{2.69} & \perfdrop{3.00} & \perfsame{0.00}
& \perfdrop{1.77} & \perfdrop{1.86} & \perfgain{0.21}
& \perfdrop{1.68} & \perfdrop{1.76} & \perfgain{0.20}
& \perfdrop{4.64} & \perfdrop{5.38} & \perfdrop{0.04}
\\

& User
& \perfgain{0.54} & \perfgain{0.27} & \perfgain{0.34}
& \perfgain{0.50} & \perfgain{0.32} & \perfgain{0.22}
& \perfgain{0.42} & \perfgain{0.18} & \perfgain{0.22}
& \perfgain{0.48} & \perfgain{0.12} & \perfgain{0.18}
& \perfgain{0.48} & \perfsame{0.00} & \perfdrop{0.05}
\\

& Time
& \perfgain{1.51} & \perfgain{1.50} & \perfgain{1.37}
& \perfgain{1.14} & \perfgain{1.13} & \perfgain{1.55}
& \perfgain{1.08} & \perfgain{1.08} & \perfdrop{0.29}
& \perfgain{1.59} & \perfgain{1.59} & \perfgain{1.43}
& \perfgain{1.20} & \perfgain{1.20} & \perfgain{1.20}
\\

\addlinespace[1pt]

\multirow{3}{*}{D3}
& CST
& \perfgain{1.98} & \perfgain{1.93} & \perfgain{1.56}
& \perfgain{2.06} & \perfgain{2.00} & \perfgain{1.48}
& \perfgain{1.47} & \perfgain{1.45} & \perfgain{0.97}
& \perfgain{1.32} & \perfgain{1.30} & \perfgain{1.11}
& \perfgain{1.23} & \perfgain{1.20} & \perfgain{0.85}
\\

& Token
& \perfdrop{9.40} & \perfdrop{9.14} & \perfdrop{6.66}
& \perfdrop{9.07} & \perfdrop{8.81} & \perfdrop{6.36}
& \perfdrop{8.15} & \perfdrop{7.80} & \perfdrop{5.97}
& \perfdrop{9.91} & \perfdrop{9.63} & \perfdrop{6.65}
& \perfdrop{8.52} & \perfdrop{8.47} & \perfdrop{4.92}
\\

& User
& \perfgain{0.91} & \perfdrop{0.94} & \perfgain{1.23}
& \perfgain{1.03} & \perfdrop{0.95} & \perfgain{1.17}
& \perfgain{1.15} & \perfdrop{0.64} & \perfgain{1.21}
& \perfdrop{0.37} & \perfdrop{1.75} & \perfgain{0.71}
& \perfdrop{0.20} & \perfdrop{1.45} & \perfgain{0.43}
\\

\midrule

\multicolumn{2}{l}{\textbf{Defect Detection}}
&
Prec. & Recall & F1
&
Prec. & Recall & F1
&
Prec. & Recall & F1
&
Prec. & Recall & F1
&
Prec. & Recall & F1
\\
\midrule

\multirow{2}{*}{D4}
& CST
& \perfdrop{5.20} & \perfgain{12.58} & \perfgain{3.25}
& \perfdrop{18.07} & \perfgain{36.58} & \perfdrop{0.22}
& \perfdrop{1.62} & \perfgain{2.31} & \perfgain{0.16}
& \perfdrop{0.12} & \perfdrop{0.32} & \perfdrop{0.21}
& \perfgain{0.58} & \perfdrop{2.47} & \perfdrop{0.55}
\\

& Token
& \perfgain{5.34} & \perfdrop{15.94} & \perfdrop{8.55}
& \perfdrop{2.49} & \perfgain{0.64} & \perfdrop{0.90}
& \perfgain{0.04} & \perfdrop{4.86} & \perfdrop{2.36}
& \perfdrop{2.03} & \perfgain{0.55} & \perfdrop{0.75}
& \perfdrop{0.02} & \perfdrop{3.98} & \perfdrop{1.50}
\\

\midrule

\multicolumn{2}{l}{\textbf{Clone Detection}}
&
Prec. & Recall & F1
&
Prec. & Recall & F1
&
Prec. & Recall & F1
&
Prec. & Recall & F1
&
Prec. & Recall & F1
\\
\midrule

\multirow{2}{*}{D5}
& CST
& \perfdrop{0.28} & \perfdrop{0.03} & \perfdrop{0.16}
& \perfgain{0.22} & \perfsame{0.00} & \perfgain{0.12}
& \perfdrop{0.37} & \perfdrop{0.32} & \perfdrop{0.35}
& \perfdrop{0.11} & \perfgain{0.01} & \perfdrop{0.05}
& \perfgain{0.09} & \perfdrop{0.03} & \perfgain{0.03}
\\

& Token
& \perfdrop{3.09} & \perfdrop{0.34} & \perfdrop{1.77}
& \perfdrop{1.65} & \perfdrop{1.28} & \perfdrop{1.48}
& \perfdrop{6.22} & \perfsame{0.00} & \perfdrop{3.33}
& \perfdrop{4.46} & \perfdrop{0.90} & \perfdrop{2.74}
& \perfdrop{10.06} & \perfgain{0.12} & \perfdrop{4.87}
\\

\midrule

\multicolumn{2}{l}{\textbf{Text-Code Retrieval}}
&
MRR & R@1 & R@5
&
MRR & R@1 & R@5
&
MRR & R@1 & R@5
&
MRR & R@1 & R@5
&
MRR & R@1 & R@5
\\
\midrule

\multirow{2}{*}{D6-Java}
& CST
& \perfgain{0.20} & \perfgain{0.40} & \perfdrop{0.06}
& \perfdrop{0.07} & \perfgain{0.07} & \perfdrop{0.17}
& \perfgain{0.01} & \perfgain{0.20} & \perfgain{0.04}
& \perfdrop{0.19} & \perfdrop{0.31} & \perfdrop{0.02}
& \perfgain{0.11} & \perfgain{0.19} & \perfgain{0.07}
\\

& Token
& \perfdrop{19.55} & \perfdrop{15.37} & \perfdrop{24.25}
& \perfdrop{14.44} & \perfdrop{11.84} & \perfdrop{16.51}
& \perfdrop{11.94} & \perfdrop{10.11} & \perfdrop{13.02}
& \perfdrop{12.86} & \perfdrop{11.32} & \perfdrop{13.83}
& \perfdrop{15.29} & \perfdrop{13.08} & \perfdrop{17.62}
\\

\addlinespace[1pt]

\multirow{2}{*}{D6-Python}
& CST
& \perfdrop{0.95} & \perfdrop{0.64} & \perfdrop{1.30}
& \perfdrop{1.18} & \perfdrop{1.24} & \perfdrop{1.15}
& \perfdrop{0.99} & \perfdrop{0.86} & \perfdrop{1.04}
& \perfdrop{1.01} & \perfdrop{1.14} & \perfdrop{0.70}
& \perfdrop{0.62} & \perfdrop{0.62} & \perfdrop{0.51}
\\

& Token
& \perfdrop{41.35} & \perfdrop{29.50} & \perfdrop{56.61}
& \perfdrop{36.27} & \perfdrop{27.61} & \perfdrop{46.98}
& \perfdrop{35.86} & \perfdrop{27.20} & \perfdrop{46.18}
& \perfdrop{30.41} & \perfdrop{24.14} & \perfdrop{37.11}
& \perfdrop{38.20} & \perfdrop{28.10} & \perfdrop{50.65}
\\

\midrule

\multicolumn{2}{l}{\textbf{Program Repair}}
&
\multicolumn{9}{c}{}
&
EM & BLEU & CodeBLEU
&
EM & BLEU & CodeBLEU
\\
\midrule

\multirow{2}{*}{D7}
& CST
& - & - & -
& - & - & -
& - & - & -
& \perfdrop{10.87} & \perfdrop{3.89} & \perfdrop{5.56}
& \perfdrop{10.56} & \perfdrop{1.70} & \perfdrop{2.02}
\\

& Token
& - & - & -
& - & - & -
& - & - & -
& \perfdrop{2.81} & \perfdrop{1.07} & \perfdrop{1.19}
& \perfdrop{5.76} & \perfdrop{3.07} & \perfdrop{2.28}
\\

\addlinespace[1pt]

\multirow{2}{*}{D8}
& CST
& - & - & -
& - & - & -
& - & - & -
& \perfdrop{7.90} & \perfdrop{0.76} & \perfdrop{1.85}
& \perfdrop{6.91} & \perfdrop{0.16} & \perfdrop{0.01}
\\

& Token
& - & - & -
& - & - & -
& - & - & -
& \perfdrop{2.08} & \perfdrop{0.85} & \perfdrop{1.05}
& \perfdrop{5.53} & \perfdrop{2.95} & \perfdrop{2.21}
\\

\midrule

\multicolumn{2}{l}{\textbf{Code Translation}}
&
\multicolumn{9}{c}{}
&
EM & BLEU & CodeBLEU
&
EM & BLEU & CodeBLEU
\\
\midrule

D9-Java$\rightarrow$C\#
& Language
& - & - & -
& - & - & -
& - & - & -
& \perfdrop{64.30} & \perfdrop{65.87} & \perfdrop{44.98}
& \perfdrop{65.30} & \perfdrop{66.48} & \perfdrop{45.66}
\\

D9-C\#$\rightarrow$Java
& Language
& - & - & -
& - & - & -
& - & - & -
& \perfdrop{67.70} & \perfdrop{61.35} & \perfdrop{50.77}
& \perfdrop{66.30} & \perfdrop{61.91} & \perfdrop{50.96}
\\

\midrule

\multicolumn{2}{l}{\textbf{API-aware Code Generation}}
&
\multicolumn{9}{c}{}
&
Syntax & Exec. & Pass@1
&
Syntax & Exec. & Pass@1
\\
\midrule

D10
& Time
& - & - & -
& - & - & -
& - & - & -
& \perfdrop{0.38} & \perfdrop{5.94} & \perfdrop{5.94}
& \perfgain{2.58} & \perfgain{2.40} & \perfgain{2.40}
\\

\bottomrule
\end{tabular}
}

\begin{tablenotes}[flushleft]
\footnotesize
\setlength{\itemsep}{0pt}
\setlength{\parskip}{0pt}
\item[] Values denote absolute numerical differences between ID and
OOD performance percentages. Red downward arrows indicate degradation, while blue upward
arrows indicate improvement.
\end{tablenotes}

\end{table*}

\section{Experimental Results}\label{sec:results}
This section presents the results for the five research questions.
All percentage-based scores are reported to two decimal places.
Absolute numerical differences between percentage-based scores
are also reported using \%, rather than as relative percentage changes.

\subsection{RQ1: Shifted Performance Evaluation}
Table~\ref{tab:rq1_performance} reports the ID--OOD performance changes across the evaluated settings. Each cell represents the performance change (i.e., the difference between the metric value of ID test and that of OOD test) for a specific setting. 

\textbf{Task-level Findings.}
Across tasks, text-code retrieval and code translation exhibit the strongest OOD degradation, with MRR decreasing by up to 41.35\% on D6-Python and Exact Match decreasing by up to 67.70\% on D9-C\#$\rightarrow$Java. In contrast, code classification and program repair are more dataset-dependent. For code classification, CST Shift leads to Accuracy decreases of 12.62--15.69\% on D2, whereas Accuracy increases by 1.23--2.06\% on D3, suggesting that structural perturbations interact strongly with dataset-specific patterns learned during fine-tuning. Text-code retrieval is particularly sensitive to Token Shift, with MRR decreasing by 30.41--41.35\% and Recall@5 decreasing by 37.11--56.61\% on D6-Python, while CST Shift causes only minor changes, possibly reflecting this task's stronger reliance on lexical correspondence. Among generation tasks, program repair shows moderate degradation, whereas code translation exhibits an Exact Match decrease of more than 64\% under Language Shift, likely because cross-language generation introduces larger lexical and syntactic differences.

\textbf{Shift-level Findings.}
Across shifts, Token and Language shifts cause the strongest degradation, User and Time shifts are generally milder, and CST Shift is highly dataset-dependent. Token Shift decreases Accuracy by 6.31--10.99\% on D1 and 8.15--9.91\% on D3. Its impact is much larger in retrieval, where MRR decreases by 11.94--19.55\% on D6-Java and 30.41--41.35\% on D6-Python. In contrast, CST Shift causes substantial degradation on D2 but improves performance on D3, showing that structural changes are not inherently harmful. Language Shift is the most severe case, causing Exact Match to decrease by more than 64\% in both translation directions, whereas User and Time shifts mostly produce minor changes. We infer that shift severity depends strongly on whether the perturbed information overlaps with task-critical signals.

\textbf{Model-level Findings.}
Across models, differences in pret-raining characteristics and architectural designs are associated with varying robustness across some settings. But these advantages do not consistently lead to stronger OOD robustness across tasks and distribution shifts. For code understanding tasks, GraphCodeBERT exhibits smaller degradation than RoBERTa and CodeBERT under CST Shift on D1, with Accuracy decreasing by 6.40\%, compared with 12.27\% and 12.07\%, respectively. This advantage may be attributed to its structure-aware pre-training objective, which incorporates code structure information beyond token sequences~\cite{guo2021graphcodebert}. However, this benefit is shift- and task-dependent rather than universal. For example, CodeT5 achieves the smallest MRR degradation under Token Shift on D6-Python (30.41\%), while CodeT5 and PLBART show different robustness patterns under other shifts despite both adopting an encoder--decoder architecture.
These results suggest that OOD robustness is not determined solely by pretraining scale or model architecture. 
Instead, it depends on the interaction between the downstream task, the type of distribution shift, and whether the learned representations capture the task-relevant information affected by that shift~\cite{quinonero2008dataset}.

\begin{rqsummarybox}
\noindent\textbf{RQ1 Summary:}
OOD degradation varies across tasks, shifts, and models. Text-code retrieval is particularly vulnerable to Token Shift (a decrease of 41.35\% in MRR), while code translation exhibits more severe degradation under Language Shift (a decrease of 67.70\% in Exact Match). No model consistently generalizes across all settings.
\end{rqsummarybox}

\begin{table*}[t]
\centering
\caption{SHIFT-WISE OOD DETECTION EFFECTIVENESS}
\label{tab:rq2_summary}

\scriptsize
\setlength{\tabcolsep}{1.8pt}
\renewcommand{\arraystretch}{1.05}

\resizebox{\textwidth}{!}{%
\begin{tabular}{
>{\centering\arraybackslash}m{1.35cm}
>{\centering\arraybackslash}m{2.15cm}
ccc ccc ccc ccc ccc
}
\toprule

\textbf{Task Family}
&
\textbf{Method}
&
\multicolumn{3}{c}{\textbf{CST Shift}}
&
\multicolumn{3}{c}{\textbf{Token Shift}}
&
\multicolumn{3}{c}{\textbf{User Shift}}
&
\multicolumn{3}{c}{\textbf{Time Shift}}
&
\multicolumn{3}{c}{\textbf{Language Shift}}
\\

\midrule

&
&
AUROC & AUPR & TNR@95
&
AUROC & AUPR & TNR@95
&
AUROC & AUPR & TNR@95
&
AUROC & AUPR & TNR@95
&
AUROC & AUPR & TNR@95
\\

\midrule

\multirow{9}{*}{\centering Classification$^{1}$}
& MSP 
&55.62&56.28&7.40
&57.12&56.68&7.80
&51.41&50.57&5.60
&50.96&50.60&5.50
&-&-&-
\\

& Entropy
&55.42&56.20&7.90
&56.94&56.50&8.40
&52.14&51.17&5.90
&51.17&50.70&5.40
&-&-&-
\\

& MaxLogit
&55.72&56.45&8.40
&56.73&56.40&8.90
&51.48&50.70&5.50
&51.00&50.60&5.50
&-&-&-
\\

& Energy
&55.21&56.07&8.20
&56.57&56.20&8.70
&51.73&50.90&5.60
&51.81&51.10&5.60
&-&-&-
\\

& Mahalanobis
&\textbf{58.50}&\textbf{58.50}&10.50
&\textbf{66.28}&\textbf{64.00}&15.30
&51.48&51.10&5.70
&55.51&54.20&6.80
&-&-&-
\\

& Mahalanobis++
&58.48&\textbf{58.50}&\textbf{10.60}
&66.27&\textbf{64.00}&\textbf{15.60}
&51.66&51.30&5.70
&55.38&53.90&6.70
&-&-&-
\\

& KNN
&57.62&57.60&8.90
&63.38&61.80&14.40
&\textbf{53.12}&\textbf{52.40}&\textbf{6.70}
&\textbf{57.41}&\textbf{55.70}&\textbf{7.90}
&-&-&-
\\

& ViM
&57.19&57.70&10.30
&58.83&58.30&12.00
&52.81&51.80&6.00
&54.71&53.30&6.30
&-&-&-
\\

& GradNorm
&53.59&54.70&7.50
&53.77&54.30&7.90
&52.06&51.30&6.00
&49.98&49.90&5.00
&-&-&-
\\

\midrule

\multirow{3}{*}{\centering Retrieval$^{2}$}
& Matching Score
&50.48&50.26&\textbf{6.06}
&\textbf{87.28}&\textbf{82.13}&\textbf{62.86}
&-&-&-
&-&-&-
&-&-&-
\\

& Ranking Entropy
&49.65&50.15&4.20
&49.28&50.00&4.04
&-&-&-
&-&-&-
&-&-&-
\\

& KNN Distance
&\textbf{50.66}&\textbf{50.48}&5.28
&58.38&54.16&13.87
&-&-&-
&-&-&-
&-&-&-
\\

\midrule

\multirow{3}{*}{\centering Generation$^{3}$}
& Seq. NLL
&\textbf{63.52}&60.06&9.57
&\textbf{56.83}&\textbf{57.64}&\textbf{7.75}
&-&-&-
&53.22&68.26&3.54
&\textbf{99.25}&\textbf{99.15}&\textbf{97.08}
\\

& Token Entropy
&42.87&45.14&4.61
&55.92&55.39&7.27
&-&-&-
&51.93&66.34&3.98
&84.71&81.72&49.00
\\

& Encoder Distance
&63.32&\textbf{62.71}&\textbf{13.47}
&54.76&55.42&6.13
&-&-&-
&\textbf{67.60}&\textbf{75.22}&\textbf{22.57}
&68.35&67.27&39.58
\\

\bottomrule
\end{tabular}%
}

\begin{tablenotes}[flushleft]
\scriptsize
\setlength{\itemsep}{0pt}
\setlength{\parskip}{0pt}

\item$^{1}$ Classification includes code classification,
defect detection, and clone detection.

\item$^{2}$ Retrieval represents text--code retrieval.

\item$^{3}$ Generation includes program repair,
code translation, and API-aware code generation.

\item[] Values are mean percentages over applicable
configurations. The best result within each task family
for each shift and metric is shown in bold.

\end{tablenotes}
\end{table*}

\begin{table*}[t]
\centering
\caption{MODEL-WISE OOD DETECTION EFFECTIVENESS}
\label{tab:rq2_model_summary}

\scriptsize
\setlength{\tabcolsep}{1.8pt}
\renewcommand{\arraystretch}{1.05}

\resizebox{\textwidth}{!}{%
\begin{tabular}{
>{\centering\arraybackslash}m{1.35cm}
>{\centering\arraybackslash}m{2.15cm}
ccc ccc ccc ccc ccc
}
\toprule

\textbf{Task Family} & \textbf{Method}
&
\multicolumn{3}{c}{\textbf{RoBERTa}}
&
\multicolumn{3}{c}{\textbf{CodeBERT}}
&
\multicolumn{3}{c}{\textbf{GraphCodeBERT}}
&
\multicolumn{3}{c}{\textbf{CodeT5}}
&
\multicolumn{3}{c}{\textbf{PLBART}}
\\

\midrule

&
&
AUROC & AUPR & TNR@95
&
AUROC & AUPR & TNR@95
&
AUROC & AUPR & TNR@95
&
AUROC & AUPR & TNR@95
&
AUROC & AUPR & TNR@95
\\

\midrule

\multirow{9}{*}{\centering Classification}

& MSP 
&51.63&51.46&5.13
&55.20&55.24&6.94
&54.64&54.64&6.04
&55.85&55.83&7.55
&57.27&56.90&9.55
\\

& Entropy
&50.91&50.82&4.81
&55.32&55.28&7.05
&55.06&55.04&6.91
&55.99&55.95&8.12
&57.50&57.23&10.38
\\

& MaxLogit
&51.38&51.38&5.86
&55.46&55.37&7.65
&54.19&54.30&6.45
&55.75&55.80&7.94
&57.40&57.14&10.76
\\

& Energy
&50.73&50.86&5.39
&55.62&55.41&7.69
&53.77&54.00&6.37
&56.16&56.01&7.98
&57.26&57.06&10.81
\\

& Mahalanobis
&\textbf{57.00}&\textbf{55.66}&\textbf{7.94}
&59.15&58.31&10.86
&59.96&58.84&10.59
&\textbf{61.88}&\textbf{60.73}&\textbf{12.95}
&59.83&59.39&12.36
\\

& Mahalanobis++
&56.98&55.65&\textbf{7.94}
&\textbf{59.29}&\textbf{58.42}&\textbf{11.06}
&\textbf{60.05}&\textbf{58.91}&\textbf{10.71}
&61.48&60.10&12.16
&\textbf{60.10}&\textbf{59.89}&\textbf{13.38}
\\

& KNN
&55.35&54.76&7.60
&59.08&58.15&10.92
&59.40&58.37&9.96
&60.43&59.21&10.98
&59.24&58.85&12.01
\\

& ViM
&51.23&51.76&6.63
&57.07&56.89&10.58
&56.99&56.42&9.11
&59.16&58.37&10.61
&58.86&58.24&11.43
\\

& GradNorm
&50.70&51.46&6.49
&53.26&53.51&7.27
&52.90&53.30&5.87
&53.52&53.98&6.48
&54.95&55.24&9.58
\\

\midrule

\multirow{3}{*}{\centering Retrieval}
& Matching Score
&\textbf{69.73}&\textbf{67.36}&\textbf{36.12}
&\textbf{69.34}&\textbf{66.42}&\textbf{34.45}
&\textbf{68.58}&\textbf{66.08}&\textbf{33.42}
&\textbf{68.39}&\textbf{65.46}&\textbf{33.32}
&\textbf{68.36}&\textbf{65.66}&\textbf{34.98}
\\

& Ranking Entropy
&50.12&50.22&4.32
&49.03&49.75&3.55
&48.54&49.86&3.50
&49.36&49.91&4.03
&50.29&50.62&5.20
\\

& KNN Distance
&56.55&53.78&10.63
&52.95&50.65&9.38
&53.47&51.63&8.62
&52.91&50.86&7.85
&56.71&54.69&11.40
\\

\midrule

\multirow{3}{*}{\centering Generation}

& Seq. NLL
&-&-&-
&-&-&-
&-&-&-
&\textbf{70.00}&\textbf{71.41}&\textbf{32.30}
&\textbf{70.70}&\textbf{72.00}&\textbf{34.08}
\\

& Token Entropy
&-&-&-
&-&-&-
&-&-&-
&58.99&60.78&17.06
&60.70&62.31&18.86
\\

& Encoder Distance
&-&-&-
&-&-&-
&-&-&-
&63.87&66.44&22.66
&61.98&61.00&17.60
\\

\bottomrule
\end{tabular}
}

\begin{tablenotes}[flushleft]
\footnotesize
\setlength{\itemsep}{0pt}
\setlength{\parskip}{0pt}
\item[] Values are mean percentages over applicable configurations. Task families follow the definitions in Table~\ref{tab:rq2_summary}. The best result for each model and metric is shown in bold.
\end{tablenotes}

\end{table*}

\subsection{RQ2: Task-adapted OOD Detection Evaluation}
Tables~\ref{tab:rq2_summary} and~\ref{tab:rq2_model_summary} summarize the effectiveness of OOD detection in three task families from shift- and model-level perspectives. The former averages results over applicable dataset--model configurations, while the latter averages over applicable dataset--shift configurations. All values are reported as mean values over the corresponding configurations.

\textbf{Shift-level Findings.}
Across shifts, OOD detectability strongly depends on both the downstream task and the selected signal. For classification tasks, Token Shift is more detectable through representation-based signals, with Mahalanobis reaching 66.28\% AUROC compared with 57.12\% for MSP, suggesting that lexical changes are more evident in the representation space than in output confidence. 
This dependence is even stronger in retrieval. Under the same Token Shift, Matching Score achieves 87.28\% AUROC, substantially outperforming Ranking Entropy (49.28\%) and KNN Distance (58.38\%).
Generation tasks exhibit a different pattern, where the target-dependent Seq. NLL achieves 99.25\% AUROC under Language Shift as an offline diagnostic, while Encoder Distance achieves the highest AUROC of 67.60\% under Time Shift.

However, classification TNR@95 remains low across methods and shifts (5.00\%--15.60\%), indicating that high AUROC or AUPR does not necessarily ensure reliable OOD rejection under strict ID retention requirements. Moreover, severe task degradation does not guarantee high detectability. For example, despite substantial performance degradation under CST Shift in RQ1, all classification detectors achieve AUROC values below 59.00\%. This suggests that OOD detectability depends on whether the selected signal captures distributional changes relevant to the task.

\textbf{Model-level Findings.}
Across models, detection effectiveness shows some backbone-specific variation, but no model provides a consistent detection advantage. 
For classification tasks, CodeT5 and PLBART generally yield higher separability than RoBERTa. For example, Mahalanobis increases from 57.00\% AUROC on RoBERTa to 61.88\% on CodeT5. 
In retrieval, however, Matching Score remains relatively consistent across all five models, with AUROC ranging from 68.36\% to 69.73\%, indicating that detection effectiveness is less sensitive to backbone choice. Ranking Entropy remains near random-level performance (48.54\%--50.29\%), while KNN Distance achieves moderate AUROC values (52.91\%--56.71\%).
For generation tasks, the relative advantage also depends on the signal: Seq.~NLL and Token Entropy perform slightly better on PLBART, whereas Encoder Distance performs better on CodeT5. 
Overall, model choice can affect how clearly a shift is exposed, but it does not determine OOD detection effectiveness by itself.

\begin{rqsummarybox}
\noindent\textbf{RQ2 Summary:}
OOD detectability is task- and signal-dependent and does not necessarily reflect performance degradation. While Matching Score achieves an AUROC of 87.28\% for retrieval under Token Shift, classification detectors remain below 59.00\% under CST Shift. The highest detection performance (AUROC reaches 99.25\%) for generation tasks relies on Seq. NLL.
\end{rqsummarybox}

\subsection{RQ3: Layer-wise Representation Diagnosis}
\begin{table*}[!t]
\centering
\caption{ID-OOD REPRESENTATION CHANGES ACROSS ANALYSIS GROUPS}
\label{tab:rq3_degraded_layerwise}

\scriptsize
\renewcommand{\arraystretch}{1.05}
\setlength{\tabcolsep}{2.0pt}

\resizebox{\textwidth}{!}{
\begin{tabular}{@{}lccccccccccc@{}}
\hline

\multirow{2}{*}{\textbf{Group}} &
\multirow{2}{*}{\textbf{$N$}} &
\multirow{2}{*}{\textbf{Alignment}} &
\multicolumn{3}{c}{\textbf{Anisotropy}} &
\multicolumn{3}{c}{\textbf{Effective Rank}} &
\multicolumn{3}{c}{\textbf{Entropy}} \\

& & &
\textbf{Mean $\Delta$ (Ratio)} &
\textbf{$p$} &
\textbf{$r$ (Mag.)} &
\textbf{Mean $\Delta$ (Ratio)} &
\textbf{$p$} &
\textbf{$r$ (Mag.)} &
\textbf{Mean $\Delta$ (Ratio)} &
\textbf{$p$} &
\textbf{$r$ (Mag.)} \\
\hline

\textit{D} Group
& 690 & 0.940
& 0.0088 (55.07) & \textbf{1.79E-21} & 0.46 (Med.)
& 11.85 (70.72) & \textbf{3.25E-36} & 0.56 (Large)
& 0.2376 (50.00) & \textbf{4.39E-11} & 0.31 (Med.) \\

\textit{ND} Group
& 512 & 0.982
& 0.0009 (41.60) & \textbf{5.66E-8} & 0.36 (Med.)
& 1.27 (50.78) & 0.18 & 0.09 (Neg.)
& 0.0644 (34.38) & 0.39 & 0.05 (Neg.) \\

\hline

Classification
& 367 & 0.959
& -0.0001 (30.25) & 0.18 & 0.12 (Small)
& 1.40 (58.04) & 3.27E-3 & 0.20 (Small)
& 0.0148 (36.78) & 0.66 & 0.03 (Neg.) \\

Retrieval
& 190 & 0.876
& 0.0281 (95.26) & \textbf{5.62E-31} & 1.00 (Large)
& 38.77 (95.26) & \textbf{5.62E-31} & 1.00 (Large)
& 0.0738 (61.58) & \textbf{1.63E-6} & 0.41 (Med.) \\

Generation
& 133 & 0.977
& 0.0057 (66.17) & \textbf{5.62E-6} & 0.47 (Med.)
& 2.22 (70.68) & 1.11E-3 & 0.33 (Med.)
& 1.0864 (69.92) & \textbf{3.98E-11} & 0.69 (Large) \\

\hline

CST
& 269 & 0.942
& 0.0041 (54.28) & \textbf{6.45E-10} & 0.48 (Med.)
& 5.78 (77.32) & \textbf{5.60E-21} & 0.68 (Large)
& 0.2755 (45.72) & 1.25E-3 & 0.24 (Small) \\

Token
& 342 & 0.930
& 0.0141 (57.60) & \textbf{8.04E-14} & 0.51 (Large)
& 22.00 (73.98) & \textbf{1.46E-30} & 0.74 (Large)
& 0.2478 (54.39) & \textbf{9.78E-10} & 0.41 (Med.) \\

\hline

RoBERTa
& 130 & 0.984
& 0.0026 (41.54) & \textbf{2.49E-5} & 0.53 (Large)
& 16.01 (68.46) & \textbf{4.34E-10} & 0.67 (Large)
& 0.0008 (43.08) & 0.12 & 0.18 (Small) \\

CodeBERT
& 104 & 0.970
& 0.0026 (53.85) & \textbf{7.44E-8} & 0.72 (Large)
& 16.20 (75.96) & \textbf{3.18E-8} & 0.66 (Large)
& 0.0138 (49.04) & 1.30E-3 & 0.42 (Med.) \\

GraphCodeBERT
& 117 & 0.972
& 0.0020 (43.59) & \textbf{2.66E-5} & 0.55 (Large)
& 15.31 (67.52) & \textbf{6.76E-8} & 0.61 (Large)
& 0.0120 (40.17) & 1.30E-2 & 0.32 (Med.) \\

CodeT5
& 234 & 0.934
& 0.0106 (62.39) & \textbf{1.66E-6} & 0.37 (Med.)
& 7.34 (71.37) & \textbf{3.63E-9} & 0.46 (Med.)
& 0.0361 (46.58) & 8.08E-3 & 0.20 (Small) \\

PLBART
& 105 & 0.829
& 0.0261 (69.52) & \textbf{9.18E-5} & 0.44 (Med.)
& 8.61 (70.48) & \textbf{1.92E-5} & 0.50 (Med.)
& 1.4527 (78.10) & \textbf{1.13E-10} & 0.74 (Large) \\

\hline
\end{tabular}
}
\begin{tablenotes}[flushleft]
\footnotesize
\setlength{\itemsep}{0pt}
\setlength{\parskip}{0pt}
\item[] $N$ denotes the number of analyzed records. 
For Anisotropy and Entropy, $\Delta=\mathrm{OOD}-\mathrm{ID}$; for Effective Rank, $\Delta=\mathrm{ID}-\mathrm{OOD}$. 
Ratio denotes the percentage of records with $\Delta>0$.
$p$ denotes the Holm-adjusted Wilcoxon signed-rank test $p$-value, and $r$ denotes the absolute rank-biserial effect size. Bold values indicate $p<0.001$.
\end{tablenotes}
\end{table*}

To examine the relationship between representation changes and task performance, we classify model--task--shift configurations as \textbf{degraded (``\textit{D}'')} only when all task performance metrics decrease from ID to OOD, and as \textbf{non-degraded (``\textit{ND}'')} which denotes the remaining configurations, based on the RQ1 results. 
Table~\ref{tab:rq3_degraded_layerwise} compares ID--OOD representation changes across different groups. We report the overall results for \textit{D} and \textit{ND} configurations, followed by task-, shift-, and model-level analysis within \textit{D}. The reported statistical tests assess paired ID and OOD differences within each group. We focus on CST and Token shifts, as other shifts contain too few \textit{D} configurations for reliable subgroup-level statistical analysis.

\textbf{\textit{D} vs. \textit{ND}.}
Compared to \textit{ND} configurations, \textit{D} configurations exhibit lower Alignment (0.940 vs. 0.982) and larger mean differences in Anisotropy (0.0088 vs. 0.0009) and Effective Rank (11.85 vs. 1.27). Effective Rank also shows a higher positive difference ratio in \textit{D} (70.72\% vs. 50.78\%). For Entropy, a significant ID--OOD difference is observed within \textit{D}, but not within \textit{ND}. These results associate OOD performance degradation with weaker cross-distribution alignment, more pronounced representation changes, and increased predictive uncertainty.

\textbf{Task-level Findings.}
Retrieval exhibits the strongest geometric changes, with Anisotropy and Effective Rank showing positive differences in 95.26\% of records (both $r=1.00$ after rounding). This indicates highly consistent directional changes, although not all records exhibit positive differences. These pronounced changes are consistent with the reliance of retrieval models on embedding similarity for ranking. Generation also exhibits pronounced changes, particularly in Entropy ($\Delta=1.0864$, $r=0.69$), suggesting greater uncertainty in token-level predictions under distribution shifts. In contrast, Classification exhibits weaker patterns, with only a small Effective Rank effect ($r=0.20$), suggesting that the strength of representation changes varies across task families.

\textbf{Shift-level Findings.}
Token exhibits larger geometric changes than CST, with higher Anisotropy and Effective Rank differences (0.0141 vs. 0.0041 and 22.00 vs. 5.78, respectively), together with lower Alignment (0.930 vs. 0.942). This suggests that changes in token distributions can substantially alter representation geometry. 
Although CST exhibits a slightly larger mean Entropy difference (0.2755 vs. 0.2478), Token shows a higher positive difference ratio and a larger effect size. These findings indicate that different shift types exhibit distinct patterns across representation diagnostics.

\textbf{Model-level Findings.}
Effective Rank provides the most consistent signal, with positive directional ratios ranging from 67.52\% to 75.96\% and statistically significant changes across all five models. This suggests that reduced OOD representation dimensionality is a relatively general characteristic of degraded configurations. In contrast, Anisotropy and Entropy exhibit greater model dependence, particularly in their directional ratios and effect sizes. 
These variations suggest that representation responses to distribution shifts are model-dependent, although differences in the evaluated task and shift settings may also contribute.

\begin{table*}[!t]
\centering
\caption{LAYER-WISE REPRESENTATION CHANGES IN \textit{D} AND \textit{ND} CONFIGURATIONS}
\label{tab:RQ3_layerwise_changes}
\scriptsize

\textbf{Panel A: 12-layer Models (RoBERTa, CodeBERT, GraphCodeBERT, and CodeT5)}
\vspace{1mm}

\begin{tabular}{@{}llc*{13}{r}@{}}
\hline

\textbf{Group} &
\textbf{$N$} &
\textbf{Metric} &
\textbf{L00} &
\textbf{L01} &
\textbf{L02} &
\textbf{L03} &
\textbf{L04} &
\textbf{L05} &
\textbf{L06} &
\textbf{L07} &
\textbf{L08} &
\textbf{L09} &
\textbf{L10} &
\textbf{L11} &
\textbf{L12} \\

\hline

\multirow{3}{*}{\textit{D}}
& \multirow{3}{*}{45}
& Anisotropy
& 0.0077 & 0.0041 & 0.0038 & 0.0040 & 0.0043 & 0.0039 & 0.0039
& 0.0053 & 0.0060 & 0.0064 & 0.0075 & 0.0076 & 0.0094 \\

& & Effective Rank
& 3.95 & 15.01 & 15.18 & 14.92 & 14.28 & 13.66 & 13.50
& 13.46 & 13.50 & 12.94 & 11.26 & 10.63 & 9.34 \\

& & Entropy
& 0.0092 & 0.0011 & -0.0024 & 0.0058 & 0.0061 & 0.0040 & 0.0045
& 0.0038 & 0.0149 & 0.0291 & 0.0423 & 0.0643 & 0.0707 \\

\hline

\multirow{3}{*}{\textit{ND}}
& \multirow{3}{*}{34}
& Anisotropy
& 0.0008 & 0.0008 & 0.0006 & 0.0007 & 0.0008 & 0.0007 & 0.0006
& 0.0007 & 0.0007 & 0.0007 & 0.0009 & 0.0017 & 0.0001 \\

& & Effective Rank
& 0.52 & 2.92 & 4.77 & 1.54 & 1.86 & 2.64 & 2.31
& 2.68 & 2.45 & 2.81 & 2.68 & 2.17 & 3.20 \\

& & Entropy
& 0.0000 & -0.0003 & -0.0004 & -0.0007 & -0.0007 & -0.0015 & -0.0021
& -0.0025 & -0.0042 & -0.0044 & -0.0049 & -0.0136 & -0.0112 \\

\hline
\end{tabular}

\vspace{1mm}

\textbf{Panel B: 6-layer Model (PLBART)}

\vspace{0.5mm}

\begin{tabular}{@{}llc*{7}{r}@{}}
\hline

\textbf{Group} &
\textbf{$N$} &
\textbf{Metric} &
\textbf{L00} &
\textbf{L01} &
\textbf{L02} &
\textbf{L03} &
\textbf{L04} &
\textbf{L05} &
\textbf{L06} \\

\hline

\multirow{3}{*}{\textit{D}}
& \multirow{3}{*}{15}
& Anisotropy
& 0.0106 & 0.0121 & 0.0116 & 0.0168 & 0.0441 & 0.0378 & 0.0495 \\

& & Effective Rank
& 18.62 & 14.92 & 11.90 & 8.78 & 3.28 & 1.85 & 0.94 \\

& & Entropy
& 2.8516 & 1.3757 & 0.4271 & 0.2391 & 0.5585 & 1.3165 & 3.4002 \\

\hline

\multirow{3}{*}{\textit{ND}}
& \multirow{3}{*}{10}
& Anisotropy
& 0.0054 & 0.0038 & 0.0050 & 0.0013 & 0.0019 & -0.0014 & -0.0029 \\

& & Effective Rank
& -9.61 & -8.04 & -7.73 & -7.31 & -4.53 & -4.77 & -3.73 \\

& & Entropy
& 0.9231 & 0.7205 & 0.3237 & 0.1527 & 0.1213 & 0.2085 & 1.0072 \\

\hline
\end{tabular}

\vspace{0.5mm}

\parbox{\textwidth}{\scriptsize
$N$ denotes the number of configurations.
Values are mean directional differences from fine-tuned representations.
For Anisotropy and Entropy, the value equals $\mathrm{OOD}-\mathrm{ID}$;
for Effective Rank, the value equals $\mathrm{ID}-\mathrm{OOD}$.
}

\end{table*}

\begin{figure*}[!ht]
    \centering
    \includegraphics[width=0.85\textwidth]{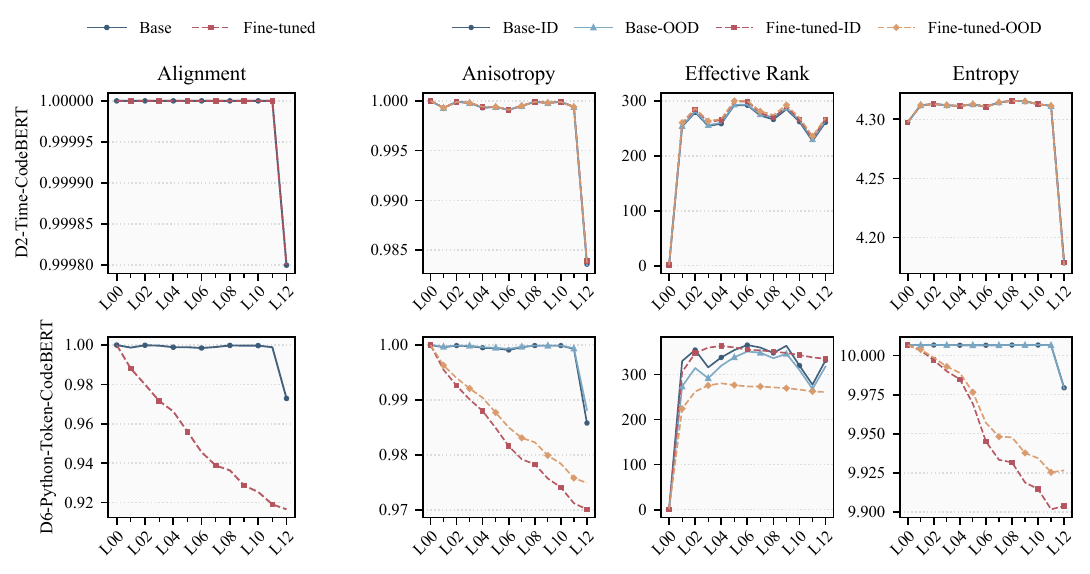}
    \caption{Layer-wise representation diagnostics of two representative CodeBERT settings before and after task-specific fine-tuning. 
    D2-Time-CodeBERT (top) represents a non-degraded configuration, while D6-Python-Token-CodeBERT (bottom) represents a degraded configuration according to RQ1.}
    \label{fig:RQ3_layerwise_diagnostics}
\end{figure*}

\textbf{Layer-wise Findings.}
Table~\ref{tab:RQ3_layerwise_changes} summarizes mean ID--OOD directional differences at each layer. We report the four 12-layer models and the 6-layer PLBART separately. 
As illustrated in the table, \textit{D} configurations show larger ID--OOD differences than \textit{ND} at nearly all layers. 
For the four 12-layer models, \textit{D} configurations exhibit larger directional changes in both Anisotropy and Effective Rank across layers. At L12, their mean differences reach 0.0094 and 9.34, respectively, compared with 0.0001 and 3.20 in \textit{ND}. This indicates stronger directional concentration and lower effective dimensionality in OOD representations of degraded configurations. Entropy shows an even clearer depth-dependent divergence. The \textit{D}-group difference increases from 0.0149 at L08 to 0.0707 at L12, whereas \textit{ND} remains negative. This suggests that geometric changes in degraded configurations are accompanied by increased predictive uncertainty in deeper layers. For PLBART, the mean ID-OOD Effective Rank difference decreases from 18.62 at L00 to 0.94 at L06, while \textit{ND} remains negative across all layers. The contrasting patterns indicate that degradation-related representation changes depend not only on network depth but also on model architecture.

Figure~\ref{fig:RQ3_layerwise_diagnostics} further compares the pretrained base and task-specific fine-tuned representations in two representative CodeBERT settings. For the \textit{ND} setting, ID and OOD diagnostic curves remain nearly overlapping before and after fine-tuning. In contrast, the \textit{D} setting exhibits more pronounced ID-OOD discrepancies after fine-tuning, particularly in deeper layers. Fine-tuned Alignment decreases to approximately 0.917 at L12, compared with roughly 0.975 for the base model. The fine-tuned model also exhibits higher OOD Anisotropy and Entropy and lower OOD Effective Rank relative to ID. These observations suggest that fine-tuning is closely relevant to greater representation divergence in the degraded setting.

\begin{rqsummarybox}
\noindent\textbf{RQ3 Summary:}
Settings with OOD performance declines exhibit weaker ID-OOD Alignment than other settings (0.940 vs. 0.982), with reduced OOD Effective Rank in 70.72\% of their layer-wise records. Representative base-finetuned comparisons further reveal greater upper-layer divergence after fine-tuning in a performance-degraded setting, suggesting an association between representation changes and OOD performance loss.
\end{rqsummarybox}

\subsection{RQ4: Traditional Adaptation under Distribution Shift}
RQ4 evaluates traditional adaptation methods (TL, EWC, and Replay) on the representative settings selected according to the results of RQ1. 
Table~\ref{tab:rq4_results} reports the ID and OOD performance before and after adaptation, with ``Baseline'' denoting the original fine-tuned models evaluated in RQ1.

\textbf{Adaptation Effectiveness.}
Traditional adaptation methods improve OOD performance in several settings, but the recoverability of a shifted model is not directly determined by the severity of its original degradation.
This is particularly evident in text--code retrieval under Token Shift.
Across the five evaluated models, adaptation increases MRR by 37.48--40.16\%, despite their substantially different baseline OOD performance.
In contrast, adaptation provides only marginal gains for classification under CST Shift. For example, on D1-CST-CodeBERT, OOD Accuracy increases from 81.52\% to at most 82.43\%.
These results indicate that a severe OOD performance drop does not necessarily imply that the corresponding shift is difficult to adapt to. 

\textbf{ID--OOD Trade-off.}
The trade-off between ID and OOD performance is also strongly shift-dependent.
For Token Shift in retrieval, adaptation improves both OOD and ID performance.
For example, on D6-Python-Token-CodeT5, EWC increases OOD MRR from 34.01\% to 71.60\% while also increasing ID MRR from 64.42\% to 81.26\%, indicating positive transfer between the two distributions.
In contrast, Language Shift in code translation introduces a pronounced conflict between OOD adaptation and ID retention.
For D9-Java$\rightarrow$C\#-CodeT5, TL and EWC improve OOD CodeBLEU from 41.91\% to 67.83\% and 67.88\%, respectively, but reduce ID CodeBLEU from 86.89\% to 52.86\% and 55.70\%.
Replay retains higher ID performance (72.17\%) at the cost of lower OOD performance (61.57\%). These results indicate that the balance between OOD adaptation and ID retention is strongly shift-dependent.

\textbf{Method-comparison Findings.}
The relative effectiveness of TL, EWC, and Replay depends on the degree of conflict between the ID and OOD distributions.
Under Token Shift in retrieval, the three methods yield nearly identical performance, suggesting that continual-learning mechanisms provide little additional benefit over standard transfer learning in this setting.
Under the more severe Language Shift, however, their differences become pronounced. TL and EWC favor OOD adaptation but suffer substantial ID forgetting, whereas Replay provides better ID retention with lower OOD performance. These results highlight the importance of considering both ID and OOD performance when selecting adaptation strategies.

\begin{rqsummarybox}
\noindent\textbf{RQ4 Summary:}
Traditional adaptation methods improve OOD performance to varying degrees, but may compromise ID retention. TL, EWC, and Replay perform similarly under Token Shift in retrieval, whereas Replay better preserves ID performance under Language Shift in translation, but at the cost of smaller OOD improvements.
\end{rqsummarybox}

\begin{table}[t]
\centering
\caption{ID AND OOD PERFORMANCE OF TRADITIONAL ADAPTATION METHODS}
\label{tab:rq4_results}

\scriptsize                     
\setlength{\tabcolsep}{2.5pt}  
\renewcommand{\arraystretch}{0.8}

\begin{tabular}{
  >{\raggedright\arraybackslash}p{1.8cm}   
  >{\centering\arraybackslash}p{0.6cm}    
  rrrrrrrr                                
}
\toprule
\textbf{Setting} &
\textbf{Metric} &
\multicolumn{2}{c}{\textbf{Baseline}} &
\multicolumn{2}{c}{\textbf{TL}} &
\multicolumn{2}{c}{\textbf{EWC}} &
\multicolumn{2}{c}{\textbf{Replay}} \\
\cmidrule(lr){3-4} \cmidrule(lr){5-6} \cmidrule(lr){7-8} \cmidrule(lr){9-10}
 & & ID & OOD & ID & OOD & ID & OOD & ID & OOD \\
\midrule

\multirow{3}{=}{D1-CST-CodeBERT} 
& Acc   & 93.59 & 81.52 & 93.51 & 81.97 & 93.55 & 82.07 & \textbf{93.65} & \textbf{82.43} \\
& M-F1  & 93.63 & 82.14 & 93.55 & 82.23 & 93.60 & 82.33 & \textbf{93.71} & \textbf{82.63} \\
& Top-5 & 96.44 & 86.18 & 96.58 & 86.77 & \textbf{96.66} & 86.77 & 96.64 & \textbf{86.93} \\
\addlinespace

\multirow{3}{=}{D1-Token-CodeBERT} 
& Acc   & 92.99 & 82.61 & 92.61 & \textbf{85.87} & 92.65 & 84.60 & \textbf{92.87} & 84.27 \\
& M-F1  & 93.06 & 82.86 & 92.73 & \textbf{86.20} & 92.73 & 84.01 & \textbf{92.94} & 83.77 \\
& Top-5 & 95.78 & 88.66 & 95.68 & \textbf{90.03} & 95.73 & 89.73 & \textbf{95.78} & 89.80 \\
\addlinespace

\multirow{3}{=}{D6-Python-Token-RoBERTa} 
& MRR  & 59.57 & 18.22 & 74.56 & \textbf{57.43} & 74.54 & 57.42 & \textbf{74.63} & 57.40 \\
& R@1  & 37.91 & 8.41 & 64.81 & \textbf{46.73} & 64.71 & 46.62 & \textbf{64.84} & 46.55 \\
& R@5  & 84.75 & 28.14 & 86.67 & 69.66 & \textbf{86.76} & \textbf{69.97} & 86.75 & 69.88 \\
\addlinespace

\multirow{3}{=}{D6-Python-Token-CodeBERT} 
& MRR  & 62.85 & 26.58 & \textbf{78.64} & 65.22 & 78.57 & 65.46 & \textbf{78.64} & \textbf{65.51} \\
& R@1  & 40.91 & 13.30 & \textbf{69.76} & 54.69 & 69.66 & 55.14 & \textbf{69.76} & \textbf{55.16} \\
& R@5  & 88.20 & 41.22 & 89.55 & \textbf{78.11} & 89.61 & 77.86 & \textbf{89.61} & 77.82 \\
\addlinespace

\multirow{3}{=}{D6-Python-Token-GraphCodeBERT} 
& MRR  & 63.75 & 27.89 & 79.84 & 67.88 & 79.84 & 67.89 & \textbf{79.86} & \textbf{68.05} \\
& R@1  & 41.52 & 14.32 & 71.14 & 57.71 & 71.14 & 57.71 & \textbf{71.16} & \textbf{58.03} \\
& R@5  & 89.08 & 42.90 & 90.64 & 80.41 & 90.66 & \textbf{80.43} & \textbf{90.71} & 80.39 \\
\addlinespace

\multirow{3}{=}{D6-Python-Token-CodeT5} 
& MRR  & 64.42 & 34.01 & 81.25 & 71.56 & \textbf{81.26} & \textbf{71.60} & 81.22 & 71.49 \\
& R@1  & 42.42 & 18.28 & \textbf{73.25} & \textbf{62.22} & 73.21 & \textbf{62.22} & 73.18 & 62.06 \\
& R@5  & 89.53 & 52.42 & 91.23 & 82.96 & 91.21 & \textbf{83.00} & \textbf{91.24} & 82.94 \\
\addlinespace

\multirow{3}{=}{D6-Python-Token-PLBART} 
& MRR  & 60.41 & 22.21 & \textbf{76.53} & \textbf{61.66} & 76.43 & 61.48 & \textbf{76.53} & \textbf{61.73} \\
& R@1  & 38.63 & 10.53 & \textbf{67.07} & 50.92 & 66.94 & 50.70 & 67.02 & \textbf{50.99} \\
& R@5  & 85.67 & 35.02 & \textbf{88.25} & \textbf{74.64} & 88.12 & 74.28 & 88.21 & 74.64 \\
\addlinespace

\multirow{3}{=}{D7-CST-CodeT5} 
& EM    & 21.90 & 11.04 & 24.80 & 22.79 & \textbf{27.88} & 22.88 & 27.54 & \textbf{23.31} \\
& BLEU  & 75.59 & 71.70 & 75.79 & \textbf{74.93} & 75.88 & 74.50 & \textbf{77.93} & 74.42 \\
& CB    & 73.27 & 67.72 & 73.59 & \textbf{72.74} & 73.82 & 72.26 & \textbf{74.80} & 72.23 \\
\addlinespace

\multirow{3}{=}{D7-CST-PLBART} 
& EM    & 18.73 & 8.17 & 19.79 & \textbf{17.57} & \textbf{20.60} & 16.88 & 19.67 & \textbf{18.51} \\
& BLEU  & 73.41 & 71.71 & 72.73 & \textbf{70.94} & 70.96 & 68.17 & \textbf{72.86} & \textbf{71.04} \\
& CB    & 71.03 & 69.01 & 70.49 & 68.87 & 68.82 & 66.00 & \textbf{70.59} & 68.97 \\
\addlinespace

\multirow{3}{=}{D9-Java$\rightarrow$C\#-CodeT5} 
& EM    & 64.30 & 0.00 & 0.00 & \textbf{41.00} & 0.00 & \textbf{41.00} & \textbf{35.20} & 34.50 \\
& BLEU  & 83.30 & 17.43 & 34.27 & 57.37 & 35.51 & \textbf{57.51} & \textbf{57.83} & 47.28 \\
& CB    & 86.89 & 41.91 & 52.86 & 67.83 & 55.70 & \textbf{67.88} & \textbf{72.17} & 61.57 \\
\addlinespace

\multirow{3}{=}{D9-Java$\rightarrow$C\#-PLBART} 
& EM    & 65.30 & 0.00 & 0.00 & \textbf{42.00} & 0.00 & \textbf{42.00} & \textbf{46.80} & 36.00 \\
& BLEU  & 84.11 & 17.63 & 20.79 & \textbf{54.37} & 20.79 & 53.94 & \textbf{67.06} & 47.71 \\
& CB    & 87.68 & 42.02 & 36.91 & \textbf{64.88} & 36.94 & 64.13 & \textbf{76.79} & 60.91 \\
\addlinespace

\multirow{3}{=}{D9-C\#$\rightarrow$Java-CodeT5} 
& EM    & 67.70 & 0.00 & 0.20 & \textbf{34.00} & 0.20 & 33.00 & \textbf{48.60} & 27.50 \\
& BLEU  & 79.69 & 18.34 & 25.83 & 56.71 & 25.77 & \textbf{57.45} & \textbf{64.16} & 49.60 \\
& CB    & 84.94 & 34.17 & 50.70 & 66.31 & 50.97 & \textbf{66.75} & \textbf{74.71} & 61.82 \\
\addlinespace

\multirow{3}{=}{D9-C\#$\rightarrow$Java-PLBART} 
& EM    & 66.30 & 0.00 & 0.10 & 24.50 & 0.10 & \textbf{26.00} & \textbf{44.00} & 25.50 \\
& BLEU  & 79.80 & 17.89 & 19.07 & \textbf{49.86} & 19.00 & 49.80 & \textbf{57.60} & 47.18 \\
& CB    & 84.90 & 33.94 & 41.17 & \textbf{58.18} & 40.64 & 58.09 & \textbf{68.11} & 57.68 \\

\bottomrule

\end{tabular}

\begin{tablenotes}[flushleft]
\footnotesize
\item[] D1, D6, D7, and D9 denote code classification,
text--code retrieval, program repair, and code translation
settings, respectively. Values are reported as percentages.
\end{tablenotes}
\end{table}

\begin{figure*}[!t]
    \centering
    \includegraphics[width=0.9\textwidth,
        keepaspectratio]{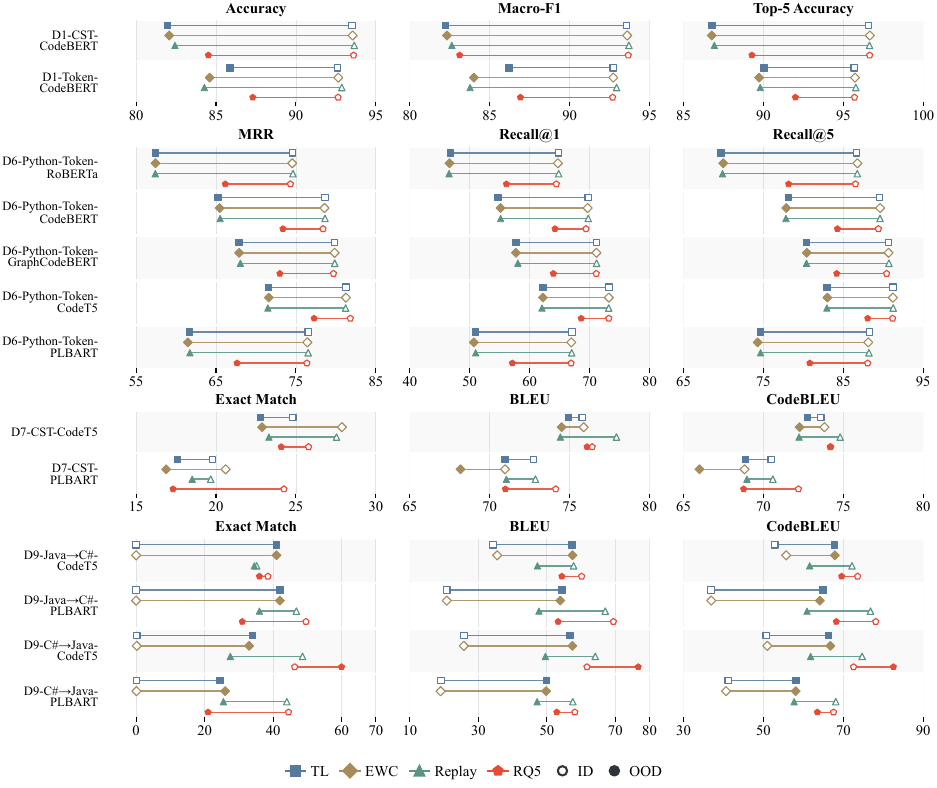}
    \caption{ID and OOD performance of traditional and representation-diagnostics-guided adaptation methods across representative distribution-shift settings. 
    Open and filled markers indicate ID and OOD performance, respectively.}
    \label{fig:rq45_adaptation}
\end{figure*}

\subsection{RQ5: Representation-guided Adaptation Selection}
RQ5 examines whether representation diagnostics can guide OOD data allocation and configuration search for Replay-based adaptation. Figure~\ref{fig:rq45_adaptation} compares the configurations selected by RQ5 with TL, EWC, and standard Replay under the representative settings used in RQ4. 

\textbf{Advantages over Replay.} The selected configurations achieve higher reported OOD validation scores than standard Replay in most settings, with particularly pronounced gains in text--code retrieval and selected translation tasks. On D6-Python-Token, OOD MRR for RoBERTa rises from 57.40\% with Replay to 66.16\% under RQ5, while ID MRR remains comparable (74.63\% vs. 74.31\%). CodeBERT similarly improves from 65.51\% to 73.37\% in OOD MRR. These results show that diagnostics-guided configuration search can identify promising Replay configurations without necessarily causing substantial additional ID degradation.

\textbf{Cross-model Consistency.} Under Token Shift in retrieval, RQ5 achieves higher reported OOD scores than standard Replay across all five models and all three metrics (MRR, Recall@1, and Recall@5), indicating a consistent pattern beyond any single model or metric. The gains are less uniform in other tasks. On D7-CST-CodeT5, OOD Exact Match increases from 23.31\% to 24.08\%, while ID Exact Match decreases from 27.54\% to 25.79\%; on D7-CST-PLBART, OOD Exact Match decreases from 18.51\% to 17.31\%. Translation results also vary by model, with improvements generally more pronounced for CodeT5 than for PLBART. Thus, the observed benefits remain task- and model-dependent.

\textbf{Configuration Transfer.} We examine surrogate-based configuration transfer in the compatible D1-Token-CodeBERT setting. Using historical configuration--performance observations to recommend candidate configurations, the selected configuration improves OOD Accuracy from 84.27\% to 87.30\% compared with standard Replay, while ID Accuracy changes from 92.87\% to 92.64\%. OOD Top-5 Accuracy also increases from 89.80\% to 92.00\%. This case provides preliminary evidence that historical search experience can support configuration selection in compatible settings.

\begin{rqsummarybox}
\noindent\textbf{RQ5 Summary:}
Diagnostics-guided configuration search achieves higher OOD validation scores than standard Replay across all retrieval models and three evaluation metrics. Additional improvements are observed in selected translation settings, although the benefits vary across tasks and models. Configuration transfer also shows preliminary promise in a compatible setting.
\end{rqsummarybox}

\section{Discussion}\label{sec:discussion}
\subsection{Beyond Performance-based Robustness Evaluation}
CodeS~\cite{hu2023codes} and CodeS+~\cite{shi2026codesplus} established important code-specific shift settings and examined code-model robustness, including detection and mitigation. Our findings extend prior evaluations across classification, retrieval, and generation tasks, revealing that OOD robustness depends on the interaction among tasks, shifts, datasets, and models. For example, Token Shift reduces retrieval MRR by 30.41-41.35\% on D6-Python, while CST Shift decreases classification Accuracy by 12.62-15.69\% on D2 but improves it by 1.23-2.06\% on D3. Moreover, no model consistently outperforms others across all settings. These findings indicate that neither shift type nor model architecture alone reliably predicts OOD robustness.

Beyond performance degradation, OOD detectability exhibits distinct task- and signal-dependent patterns. Despite substantial classification degradation under CST Shift, all evaluated detectors achieve AUROC values below 59.00\%, whereas Matching Score reaches 87.28\% under Token Shift in retrieval. For generation, Seq. NLL achieves high OOD detection performance but requires reference outputs, making it unsuitable for reference-free deployment. These findings suggest that performance degradation alone is insufficient to characterize OOD reliability. Detection effectiveness and adaptation outcomes must also be considered, as severe shifts are not necessarily easier to detect or harder to mitigate.

\subsection{Representation Diagnostics and Adaptation Implications}
RQ3 reveals that OOD performance degradation is closely associated with changes in representation alignment, geometry, and predictive uncertainty. Moreover, these changes vary across network layers and model architectures. The base-finetuned comparison further shows more pronounced ID-OOD divergence after fine-tuning in a performance-degraded setting, but not in a non-degraded setting. These findings highlight the potential value of layer-aware representation diagnostics for understanding OOD failures.

The adaptation results further highlight the need to balance OOD improvement against ID retention. For example, in D9-Java$\rightarrow$C\#-CodeT5, Replay preserves higher ID CodeBLEU than TL (72.17\% vs. 52.86\%) but achieves lower OOD CodeBLEU (61.57\% vs. 67.83\%). Building on these observations, RQ5 shows that diagnostics-guided configuration search achieves higher reported OOD validation scores than standard Replay across five retrieval models and three metrics, including an MRR improvement from 57.40\% to 66.16\% for RoBERTa. Together, these findings suggest that representation diagnostics can support adaptation configuration selection, while effective adaptation must also account for ID-OOD trade-offs.

\subsection{Future Directions for Representation-Aware Adaptation}
Our findings suggest two directions for future research:
\begin{itemize}
    \item \textbf{Richer Representation Diagnostics:}
    The four diagnostics used in this study primarily capture global geometry and predictive uncertainty. Future studies could examine complementary local and structural properties using CKA~\cite{kornblith2019similarity}, RSA~\cite{kriegeskorte2008representational}, Local Intrinsic Dimensionality~\cite{kamkari2024lid}, and syntax- or data-flow-aware representations~\cite{allamanis2018learning,guo2021graphcodebert}. Such signals could help distinguish distribution changes that affect task-relevant code semantics from changes that leave predictions largely unaffected.

    \item \textbf{Reliable Layer-aware Adaptation:}
    Because representation changes are not uniform across depth, future adaptation methods could combine information from selected intermediate layers with layer-specific parameter updates or regularization. This would extend the current final-layer-guided configuration strategy without assuming that deeper layers always exhibit stronger degradation.
\end{itemize}

\section{Threats to Validity}\label{sec:threats}
\textbf{Internal Validity.} We use official model checkpoints, established evaluation protocols, and a consistent experimental environment to minimize implementation-related variations. Nevertheless, NatGen and Tree-sitter transformations cannot guarantee semantic equivalence for every transformed sample, potentially affecting the validity of constructed shifts. In addition, data sampling and stochastic optimization may introduce experimental variability. Another potential threat lies in the configuration search of RQ5, where OOD evaluation data are repeatedly used for TPE search and Pareto selection. Moreover, the class composition of these evaluation subsets may vary across trials, affecting the comparability of configurations and potentially introducing selection bias. Consequently, the reported adaptation gains may be optimistic.

\textbf{Construct Validity.} Task-specific performance metrics and four representation diagnostics provide complementary perspectives on OOD robustness, but cannot fully capture software reliability or code semantics. In particular, centroid-based Alignment under non-paired shifts may be influenced by class distributions, while Anisotropy and Effective Rank characterize global representation geometry and may overlook local structural variations. Entropy, in contrast, reflects predictive uncertainty rather than representation geometry directly. Furthermore, the degradation criterion requires declines in all three task metrics, potentially excluding other meaningful failure patterns and influencing the distinction between degraded and non-degraded settings. Layer-wise observations from the same configuration are also correlated, which may affect the interpretation of aggregate statistical significance.

\textbf{External Validity.} Our study covers five code models, seven SE tasks, and five distribution shift types, but excludes large-scale decoder-only models and has uneven dataset and architecture coverage across tasks. Moreover, analyzed code shifts may not fully reflect real-world software evolution, where multiple distributional factors can interact or evolve over time. The adaptation experiments also focus on representative challenging settings rather than all evaluated configurations, potentially limiting the generalizability of their findings. Furthermore, these experiments assume access to labeled OOD samples and retained ID data, which may not always be available in practical deployment. Thus, the effectiveness of the adaptation methods under more diverse and constrained deployment conditions remains to be established.

\section{Conclusions}\label{sec:conclusion}
This paper presents a systematic empirical study of code representation models under software distribution shifts, covering performance degradation, OOD detection, representation changes, and adaptation across seven SE tasks, five models, and five shift types. Our findings reveal that OOD degradation varies across tasks, shifts, and models. Text-code retrieval is particularly vulnerable to Token Shift (a decrease of up to 41.35\% in MRR), while code translation suffers more severe degradation under Language Shift (a decrease of up to 67.70\% in Exact Match). Meanwhile, severe performance degradation does not necessarily imply high OOD detectability. Representation analysis further identifies associations between performance degradation and layer-dependent changes in alignment, geometry, and predictive uncertainty. 
Moreover, traditional adaptation methods exhibit varying ID-OOD trade-offs, whereas diagnostics-guided configuration search yields
higher OOD validation scores than standard Replay in all retrieval models, with additional gains in selected translation settings. 

These results have practical implications for evaluating and adapting code models. Robustness assessment should extend beyond task performance to consider OOD detectability and representation behavior, while adaptation decisions should account for both OOD improvement and ID retention rather than prioritizing either in isolation. Future work will explore richer multi-layer diagnostics and more reliable adaptation strategies, extending the investigation to larger code models and real-world distribution shifts.


\begin{thebibliography}{99}

\bibitem{liu2019RoBERTa}
Y. Liu et al.,
``RoBERTa: A robustly optimized BERT pretraining approach,''
2019, arXiv:1907.11692.

\bibitem{feng2020codebert}
Z. Feng et al.,
``CodeBERT: A pre-trained model for programming and natural languages,''
in \textit{Findings Assoc. Comput. Linguistics: EMNLP}, 2020,
pp. 1536--1547.

\bibitem{guo2021graphcodebert}
D. Guo et al.,
``GraphCodeBERT: Pre-training code representations with data flow,''
in \textit{Proc. Int. Conf. Learn. Represent.}, 2021.

\bibitem{wang2021codet5}
Y. Wang, W. Wang, S. Joty, and S. C. H. Hoi,
``CodeT5: Identifier-aware unified pre-trained encoder-decoder models
for code understanding and generation,''
in \textit{Proc. Conf. Empirical Methods Natural Language Process.},
2021, pp. 8696--8708.

\bibitem{hu2023codes}
Q. Hu, Y. Guo, X. Xie, M. Cordy, M. Papadakis, L. Ma, and Y. Le Traon,
``CodeS: Towards code model generalization under distribution shift,''
in \textit{Proc. IEEE/ACM Int. Conf. Softw. Eng. New Ideas Emerg. Results},
2023, pp. 1--6.

\bibitem{shi2026codesplus}
Z. Shi et al.,
``CodeS+: Towards assessing the generalization ability of code models
under distribution shift,''
\textit{IEEE Trans. Softw. Eng.}, vol. 52, no. 4, pp. 1515--1530,
Apr. 2026.

\bibitem{hendrycks2017baseline}
D. Hendrycks and K. Gimpel,
``A baseline for detecting misclassified and out-of-distribution examples
in neural networks,''
in \textit{Proc. Int. Conf. Learn. Represent.}, 2017.

\bibitem{lee2018mahalanobis}
K. Lee, K. Lee, H. Lee, and J. Shin,
``A simple unified framework for detecting out-of-distribution samples
and adversarial attacks,''
in \textit{Proc. Adv. Neural Inf. Process. Syst.}, vol. 31, 2018,
pp. 7167--7177.

\bibitem{liu2020energy}
W. Liu, X. Wang, J. D. Owens, and Y. Li,
``Energy-based out-of-distribution detection,''
in \textit{Proc. Adv. Neural Inf. Process. Syst.}, vol. 33, 2020,
pp. 21464--21475.

\bibitem{wang2020understanding}
T. Wang and P. Isola,
``Understanding contrastive representation learning through alignment
and uniformity on the hypersphere,''
in \textit{Proc. Int. Conf. Mach. Learn.}, 2020, pp. 9929--9939.

\bibitem{ethayarajh2019contextual}
K. Ethayarajh,
``How contextual are contextualized word representations? Comparing the
geometry of BERT, ELMo, and GPT-2 embeddings,''
in \textit{Proc. Conf. Empirical Methods Natural Language Process.
Int. Joint Conf. Natural Language Process.}, 2019, pp. 55--65.

\bibitem{roy2007effective}
O. Roy and M. Vetterli,
``The effective rank: A measure of effective dimensionality,''
in \textit{Proc. Eur. Signal Process. Conf.}, 2007, pp. 606--610.

\bibitem{kirsch2021pitfalls}
A. Kirsch, J. Mukhoti, J. van Amersfoort, P. H. S. Torr, and Y. Gal,
``On pitfalls in out-of-distribution detection: Predictive entropy
considered harmful,''
in \textit{Proc. ICML Workshop Uncertainty Robustness Deep Learning},
2021.

\bibitem{pan2010survey}
S. J. Pan and Q. Yang,
``A survey on transfer learning,''
\textit{IEEE Trans. Knowl. Data Eng.}, vol. 22, no. 10,
pp. 1345--1359, Oct. 2010.

\bibitem{parisi2019continual}
G. I. Parisi, R. Kemker, J. L. Part, C. Kanan, and S. Wermter,
``Continual lifelong learning with neural networks: A review,''
\textit{Neural Netw.}, vol. 113, pp. 54--71, 2019.

\bibitem{delange2022continual}
M. De Lange et al.,
``A continual learning survey: Defying forgetting in classification tasks,''
\textit{IEEE Trans. Pattern Anal. Mach. Intell.}, vol. 44, no. 7,
pp. 3366--3385, Jul. 2022.

\bibitem{robins1995catastrophic}
A. Robins,
``Catastrophic forgetting, rehearsal and pseudorehearsal,''
\textit{Connection Sci.}, vol. 7, no. 2, pp. 123--146, 1995.

\bibitem{allamanis2018learning}
M. Allamanis, M. Brockschmidt, and M. Khademi,
``Learning to represent programs with graphs,''
in \textit{Proc. Int. Conf. Learn. Represent.}, 2018.

\bibitem{alon2019code2vec}
U. Alon, M. Zilberstein, O. Levy, and E. Yahav,
``code2vec: Learning distributed representations of code,''
\textit{Proc. ACM Program. Lang.}, vol. 3, no. POPL,
Art. no. 40, pp. 40:1--40:29, Jan. 2019.

\bibitem{alon2019code2seq}
U. Alon et al.,
``code2seq: Generating sequences from structured representations of code,''
in \textit{Proc. Int. Conf. Learn. Represent.}, 2019.

\bibitem{ahmad2021unified}
W. U. Ahmad, S. Chakraborty, B. Ray, and K.-W. Chang,
``Unified pre-training for program understanding and generation,''
in \textit{Proc. Conf. North Amer. Chapter Assoc. Comput. Linguistics:
Human Language Technol.}, 2021, pp. 2655--2668.

\bibitem{chen2021evaluating}
M. Chen et al.,
``Evaluating large language models trained on code,''
2021, arXiv:2107.03374.

\bibitem{quinonero2008dataset}
J. Qui\~nonero-Candela, M. Sugiyama, A. Schwaighofer, and N. D. Lawrence,
eds.,
\textit{Dataset Shift in Machine Learning}.
Cambridge, MA, USA: MIT Press, 2008.

\bibitem{koh2021wilds}
P. W. Koh et al.,
``WILDS: A benchmark of in-the-wild distribution shifts,''
in \textit{Proc. Int. Conf. Mach. Learn.}, vol. 139, 2021,
pp. 5637--5664.

\bibitem{yang2023generalized}
J. Yang, K. Zhou, Y. Li, and Z. Liu,
``Generalized out-of-distribution detection: A survey,''
\textit{Int. J. Comput. Vis.}, vol. 132, pp. 5635--5662, 2024.

\bibitem{husain2019codesearchnet}
H. Husain, H.-H. Wu, T. Gazit, M. Allamanis, and M. Brockschmidt,
``CodeSearchNet challenge: Evaluating the state of semantic code search,''
2019, arXiv:1909.09436.

\bibitem{lu2021codexglue}
S. Lu et al.,
``CodeXGLUE: A machine learning benchmark dataset for code understanding
and generation,''
in \textit{Proc. Adv. Neural Inf. Process. Syst. Datasets Benchmarks Track},
2021.

\bibitem{puri2021codenet}
R. Puri et al.,
``CodeNet: A large-scale AI for code dataset for learning a diversity
of coding tasks,''
in \textit{Proc. Adv. Neural Inf. Process. Syst. Datasets Benchmarks Track},
2021.

\bibitem{niu2023crosscodebench}
C. Niu, C. Li, V. Ng, and B. Luo,
``CrossCodeBench: Benchmarking cross-task generalization of source code
models,''
in \textit{Proc. IEEE/ACM Int. Conf. Softw. Eng.}, 2023.

\bibitem{niu2023empiricalcomparison}
C. Niu, C. Li, V. Ng, D. Chen, J. Ge, and B. Luo,
``An empirical comparison of pre-trained models of source code,''
in \textit{Proc. IEEE/ACM Int. Conf. Softw. Eng.}, 2023,
pp. 2136--2148.

\bibitem{arakelyan2023distributional}
S. Arakelyan, R. Das, Y. Mao, and X. Ren,
``Exploring distributional shifts in large language models for code
analysis,''
in \textit{Proc. Conf. Empirical Methods Natural Language Process.},
2023, pp. 16298--16314.

\bibitem{weyssow2023continual}
M. Weyssow, X. Zhou, K. Kim, D. Lo, and H. Sahraoui,
``On the usage of continual learning for out-of-distribution generalization
in pre-trained language models of code,''
in \textit{Proc. ACM Joint Eur. Softw. Eng. Conf. Symp. Found. Softw. Eng.},
2023, pp. 1470--1482.

\bibitem{liu2024codeupdatearena}
Z. L. Liu, S. Pandit, X. Ye, E. Choi, and G. Durrett,
``CodeUpdateArena: Benchmarking knowledge editing on API updates,''
2024, arXiv:2407.06249.

\bibitem{papineni2002bleu}
K. Papineni, S. Roukos, T. Ward, and W.-J. Zhu,
``BLEU: A method for automatic evaluation of machine translation,''
in \textit{Proc. Annu. Meeting Assoc. Comput. Linguistics}, 2002,
pp. 311--318.

\bibitem{ren2020codebleu}
S. Ren et al.,
``CodeBLEU: A method for automatic evaluation of code synthesis,''
2020, arXiv:2009.10297.

\bibitem{hendrycks2022scaling}
D. Hendrycks, S. Basart, M. Mazeika, et al.,
``Scaling out-of-distribution detection for real-world settings,''
in \textit{Proc. Int. Conf. Mach. Learn.}, 2022, pp. 8759--8773.

\bibitem{muller2025mahalanobis}
M. M\"uller and M. Hein,
``Mahalanobis++: Improving OOD detection via feature normalization,''
in \textit{Proc. Int. Conf. Mach. Learn.}, vol. 267, 2025,
pp. 45151--45184.

\bibitem{sun2022knn}
Y. Sun, Y. Ming, X. Zhu, and Y. Li,
``Out-of-distribution detection with deep nearest neighbors,''
in \textit{Proc. Int. Conf. Mach. Learn.}, vol. 162, 2022,
pp. 20827--20840.

\bibitem{wang2022vim}
H. Wang, Z. Li, L. Feng, and W. Zhang,
``ViM: Out-of-distribution with virtual-logit matching,''
in \textit{Proc. IEEE/CVF Conf. Comput. Vis. Pattern Recognit.},
2022, pp. 4921--4930.

\bibitem{huang2021gradnorm}
R. Huang, A. Geng, and Y. Li,
``On the importance of gradients for detecting distributional shifts
in the wild,''
in \textit{Proc. Adv. Neural Inf. Process. Syst.}, vol. 34, 2021,
pp. 677--689.

\bibitem{hajipour2024simscood}
H. Hajipour, N. Yu, C.-A. Staicu, and M. Fritz,
``SimSCOOD: Systematic analysis of out-of-distribution generalization
in fine-tuned source code models,''
in \textit{Findings Assoc. Comput. Linguistics: NAACL},
2024, pp. 1400--1416.

\bibitem{kirkpatrick2017ewc}
J. Kirkpatrick et al.,
``Overcoming catastrophic forgetting in neural networks,''
\textit{Proc. Natl. Acad. Sci. USA}, vol. 114, no. 13,
pp. 3521--3526, Mar. 2017.

\bibitem{chaudhry2019episodic}
A. Chaudhry et al.,
``On tiny episodic memories in continual learning,''
2019, arXiv:1902.10486.

\bibitem{bergstra2011tpe}
J. Bergstra, R. Bardenet, Y. Bengio, and B. K\'egl,
``Algorithms for hyper-parameter optimization,''
in \textit{Proc. Adv. Neural Inf. Process. Syst.}, vol. 24, 2011,
pp. 2546--2554.

\bibitem{chakraborty2022natgen}
S. Chakraborty, T. Ahmed, Y. Ding, P. T. Devanbu, and B. Ray,
``NatGen: Generative pre-training by `naturalizing' source code,''
in \textit{Proc. ACM Joint Eur. Softw. Eng. Conf. Symp. Found. Softw. Eng.},
2022, pp. 18--30.

\bibitem{treesitter}
Tree-sitter Contributors,
``Tree-sitter: An incremental parsing system for programming tools,''
[Online]. Available: \url{https://github.com/tree-sitter/tree-sitter}.
Accessed: Jul. 15, 2026.

\bibitem{deb2002nsga2}
K. Deb, A. Pratap, S. Agarwal, and T. Meyarivan,
``A fast and elitist multiobjective genetic algorithm: NSGA-II,''
\textit{IEEE Trans. Evol. Comput.}, vol. 6, no. 2,
pp. 182--197, Apr. 2002.

\bibitem{karmakar2021codeknowledge}
A. Karmakar and R. Robbes,
``What do pre-trained code models know about code?''
in \textit{Proc. IEEE/ACM Int. Conf. Automated Softw. Eng.},
2021, pp. 1332--1336.

\bibitem{khan2024xcodeeval}
M. A. M. Khan, M. S. Bari, X. L. Do, W. Wang, M. R. Parvez,
and S. Joty,
``XCodeEval: An execution-based large scale multilingual multitask
benchmark for code understanding, generation, translation and retrieval,''
in \textit{Proc. Annu. Meeting Assoc. Comput. Linguistics}, 2024,
pp. 6766--6805.

\bibitem{yan2024codescope}
W. Yan et al.,
``CodeScope: An execution-based multilingual multitask multidimensional
benchmark for evaluating LLMs on code understanding and generation,''
in \textit{Proc. Annu. Meeting Assoc. Comput. Linguistics}, 2024,
pp. 5511--5558.

\bibitem{troshin2022probing}
S. Troshin and N. Chirkova,
``Probing pretrained models of source codes,''
in \textit{Proc. 5th BlackboxNLP Workshop Analyzing and Interpreting
Neural Networks for NLP}, 2022, pp. 371--383.

\bibitem{karmakar2024inspect}
A. Karmakar and R. Robbes,
``INSPECT: Intrinsic and systematic probing evaluation for code
transformers,''
\textit{IEEE Trans. Softw. Eng.}, vol. 50, no. 2,
pp. 220--238, Feb. 2024.

\bibitem{shi2023efficient}
E. Shi et al.,
``Towards efficient fine-tuning of pre-trained code models:
An experimental study and beyond,''
in \textit{Proc. ACM SIGSOFT Int. Symp. Softw. Testing Anal.},
2023, pp. 39--51.

\bibitem{niu2022survey}
C. Niu, C. Li, B. Luo, and V. Ng,
``Deep learning meets software engineering: A survey on pre-trained
models of source code,''
in \textit{Proc. Int. Joint Conf. Artif. Intell.}, 2022,
pp. 5546--5555.

\bibitem{li2021uncertainty}
Y. Li, S. Chen, and W. Yang,
``Estimating predictive uncertainty under program data distribution shift,''
2021, arXiv:2107.10989.

\bibitem{nie2022evaluation}
P. Nie, J. Zhang, J. J. Li, R. Mooney, and M. Gligoric,
``Impact of evaluation methodologies on code summarization,''
in \textit{Proc. Annu. Meeting Assoc. Comput. Linguistics}, 2022,
pp. 4936--4960.

\bibitem{zhou2019devign}
Y. Zhou, S. Liu, J. Siow, X. Du, and Y. Liu,
``Devign: Effective vulnerability identification by learning comprehensive
program semantics via graph neural networks,''
in \textit{Proc. Adv. Neural Inf. Process. Syst.}, vol. 32, 2019,
pp. 10197--10207.

\bibitem{white2016clone}
M. White, M. Tufano, C. Vendome, and D. Poshyvanyk,
``Deep learning code fragments for code clone detection,''
in \textit{Proc. IEEE/ACM Int. Conf. Automated Softw. Eng.},
2016, pp. 87--98.

\bibitem{gu2018deepcodesearch}
X. Gu, H. Zhang, and S. Kim,
``Deep code search,''
in \textit{Proc. IEEE/ACM Int. Conf. Softw. Eng.},
2018, pp. 933--944.

\bibitem{chen2021sequencer}
Z. Chen, S. Kommrusch, M. Tufano, L.-N. Pouchet,
D. Poshyvanyk, and M. Monperrus,
``SequenceR: Sequence-to-sequence learning for end-to-end program repair,''
\textit{IEEE Trans. Softw. Eng.}, vol. 47, no. 9,
pp. 1943--1959, Sep. 2021.

\bibitem{subramani2025simba}
N. Subramani, A. Gomez, and M. T. Diab,
``SimBA: Simplifying benchmark analysis using performance matrices alone,''
in \textit{Findings Assoc. Comput. Linguistics: EMNLP},
2025, pp. 13220--13233.

\bibitem{zhan2022benchsubset}
H. Zhan et al.,
``BenchSubset: A framework for selecting benchmark subsets based on
consensus clustering,''
\textit{Int. J. Intell. Syst.}, vol. 37, no. 8,
pp. 5248--5271, 2022.

\bibitem{bendavid2006neurips}
S. Ben-David, J. Blitzer, K. Crammer, and F. Pereira,
``Analysis of representations for domain adaptation,''
in \textit{Proc. Adv. Neural Inf. Process. Syst.}, vol. 19,
2006, pp. 137--144.

\bibitem{zhao2019invariant}
H. Zhao, R. Tachet des Combes, K. Zhang, and G. Gordon,
``On learning invariant representations for domain adaptation,''
in \textit{Proc. Int. Conf. Mach. Learn.}, vol. 97, 2019,
pp. 7523--7532.

\bibitem{lu2025stability}
A. Lu, H. Yuan, T. Feng, and Y. Sun,
``Rethinking the stability-plasticity trade-off in continual learning
from an architectural perspective,''
in \textit{Proc. Int. Conf. Mach. Learn.}, vol. 267, 2025,
pp. 40888--40902.

\bibitem{kornblith2019similarity}
S. Kornblith, M. Norouzi, H. Lee, and G. Hinton,
``Similarity of neural network representations revisited,''
in \textit{Proc. Int. Conf. Mach. Learn.}, vol. 97,
2019, pp. 3519--3529.

\bibitem{kriegeskorte2008representational}
N. Kriegeskorte, M. Mur, and P. Bandettini,
``Representational similarity analysis: Connecting the branches
of systems neuroscience,''
\textit{Front. Syst. Neurosci.}, vol. 2, Art. no. 4, 2008.

\bibitem{kamkari2024lid}
H. Kamkari, B. L. Ross, J. C. Cresswell, A. L. Caterini,
R. G. Krishnan, and G. Loaiza-Ganem,
``A geometric explanation of the likelihood OOD detection paradox,''
in \textit{Proc. Int. Conf. Mach. Learn.}, vol. 235,
2024, pp. 22908--22935.

\bibitem{zuo2021challenging}
L. Zuo, M. Jing, J. Li, L. Zhu, K. Lu, and Y. Yang,
``Challenging tough samples in unsupervised domain adaptation,''
\textit{Pattern Recognit.}, vol. 110, Art. no. 107540, 2021.

\bibitem{lifshitz2021sample}
O. Lifshitz and L. Wolf,
``Sample selection for universal domain adaptation,''
in \textit{Proc. AAAI Conf. Artif. Intell.},
vol. 35, no. 10, pp. 8592--8600, 2021.

\end{thebibliography}
\end{document}